%% file: main.tex
\documentclass{iopjournal}

\usepackage{lipsum}
\usepackage{subcaption}

\usepackage[backend=biber]{biblatex}
\usepackage{glossaries}
\glsdisablehyper
\loadglsentries{acronyms}

\usepackage{dsfont}
\usepackage{siunitx}
\usepackage{lmodern}

\begin{document}

\articletype{Paper} 

\title{The Impact of Physically Motivated Calibration Errors on Search Pipeline Detection Parameters for Broadband Burst Signals}

\author{
	Milan Wils$^{1,*}$\orcid{0000-0002-1544-7193},
	Brad Ratto$^{2,3}$\orcid{0009-0007-5810-1719},
	Jeffrey S. Kissel$^4$\orcid{0000-0002-1702-9577},
	Tjonnie G. F. Li$^{1,5}$\orcid{0000-0003-4297-7365},
	Marek J. Szczepańczyk$^6$\orcid{0000-0002-6167-6149},
	Gabriele Vedovato$^7$\orcid{0000-0001-7226-1320},
	and Michele Zanolin$^8$\orcid{0000-0002-4044-4306}
}

\affil{
	$^1$
	KU Leuven,
	Leuven Gravity Institute,
	Department of Physics and Astronomy,
	Celestijnenlaan 200D box 2415, 3001 Leuven, Belgium
}

\affil{
	$^2$
	Department of Mechanical and Aerospace Engineering,
	University of California San Diego,
	La Jolla, CA 92093 USA
}

\affil{
	$^3$
	Los Alamos National Laboratory,
	Los Alamos, NM 87545, USA
}

\affil{
	$^4$
	LIGO Hanford Observatory,
	Richland, WA 99352, USA
}

\affil{
	$^5$
	KU Leuven,
	Leuven Gravity Institute,
	Department of Electrical Engineering (ESAT),
	STADIUS Center for Dynamical Systems, Signal Processing and Data Analytics,
	Kasteelpark Arenberg 10, 3001 Leuven, Belgium
}

\affil{
	$^6$
	Faculty of Physics,
	University of Warsaw,
	Ludwika Pasteura 5, 02-093 Warszawa, Poland
}

\affil{
	$^7$
	INFN,
	Sezione di Padova,
	I-35131 Padova, Italy
}

\affil{
	$^8$
	Embry-Riddle Aeronautical University,
	Prescott, AZ 86301, USA
}

\affil{$^*$Author to whom any correspondence should be addressed.}

\email{milan.wils@kuleuven.be}


\date{\today}

\begin{abstract}
	Gravitational-wave observatories rely on precise calibration to convert raw detector readouts into an estimate of the gravitational-wave strain.
	Imperfections in this calibration introduce frequency dependent amplitude and phase errors on the measured gravitational wave signal.
	Previous unmodelled burst searches have approximated these effects using prescriptions such as a uniform amplitude rescaling or a constant time shift, which do not capture the frequency-dependent structure of calibration errors.
	This limitation is problematic for core-collapse supernovae, whose predicted gravitational-wave signals occupy a wide frequency band and exhibit complex time-frequency morphology.

	In this work, we investigate how realistic calibration errors affect burst search pipelines by combining analytical modelling with large-scale injection campaigns.
	First-order analytical estimates are derived to quantify how frequency-dependent amplitude and phase errors influence detection statistics such as the coherent network signal-to-noise ratio and the correlation coefficient.
	These calculations predict that the relative impact on the coherent network signal-to-noise ratio scales with the signal strength until it reaches an asymptotic value.
	The effect on the correlation coefficient is most pronounced near the detection threshold and is entirely suppressed at high signal-to-noise ratio.

	To test these predictions, we developed a plugin for the coherent WaveBurst pipeline that simulates the effect of realistic calibration errors.
	Injection studies confirm that calibration errors do modify the detection statistics, but show that the dominant contribution arises indirectly through changes in the number of time-frequency pixels selected in an event.

	Despite these measurable variations, detection efficiencies as a function of distance differ by less than one percent across all tested waveforms, and explosion-energy limits remain dominated by astrophysical uncertainties rather than calibration uncertainty.
	These results demonstrate that, at current detector sensitivity, realistic calibration errors have minimal impact on the detectability of broadband gravitational-wave burst signals.
	The impact of calibration errors on parameter estimation is left for future work.
\end{abstract}

\section{\label{sec:intro} Introduction}

The direct detection of \glspl{gw} has opened a new observational window onto the universe~\cite{2016PhRvD..93l2004A,2025arXiv250818082T,2026arXiv260527225T} and enabled multi‑messenger investigations of some of the most energetic astrophysical phenomena~\cite{2017ApJ...848L..13A}.
Ground‑based interferometric observatories in the \gls{lvk} network~\cite{2021PTEP.2021eA101A, 2015CQGra..32b4001A, 2015CQGra..32g4001L} detect gravitational waves through the length variations they introduce between test masses, and accurate reconstruction of the gravitational‑wave strain requires a precise understanding of the detector response.
This process, known as calibration, involves modelling and tracking the interferometer’s behaviour over time to obtain a faithful estimate of the dimensionless strain $h(t)$~.
Because the detector response is neither perfectly known nor perfectly stable, the reconstructed strain inevitably carries both systematic calibration errors and statistical calibration uncertainty~\cite{2017PhRvD..96j2001C,2020CQGra..37v5008S,2021arXiv210700129S,2022CQGra..39d5006A}.

The importance of accurate calibration was already recognised during the first \gls{gw} detection, GW150914, for which the amplitude and phase uncertainties were of order ten percent and ten degrees, respectively~\cite{2017PhRvD..95f2003A}.
Since then, extensive effort has led to significant improvements in calibration accuracy, uncertainty quantification, real‑time tracking of temporal variations in the response and calibration using astrophysical sources~\cite{2026arXiv260511703T}.
These improvements are essential because calibration errors affect not only parameter‑estimation analyses~\cite{2020PhRvD.102l2004P,2021PhRvD.103f3016V,2012PhRvD..85f4034V,2025PhRvD.112h4038S} but also the ability of search pipelines to identify candidate \gls{gw} events~\cite{2022PhRvD.105h2002E}.
Additionally, several studies have highlighted the role of calibration systematics in tests of general relativity~\cite{2019CQGra..36t5006H,2024arXiv240502197G}, measurements of the Hubble constant~\cite{2025PhRvD.111f3034H} and detection of the stochastic gravitational‑wave background~\cite{2023PhRvD.107j2002Y}.
However, the estimation of the impact of calibration errors on search pipelines for unmodelled \gls{gw} \emph{burst} signals has been limited to simplified approximations, and a detailed investigation of the impact of realistic calibration errors on burst detection statistics and astrophysical conclusions has not yet been performed.

For searches targeting gravitational waves from \glspl{ccsn} the situation is particularly subtle because they have a complex signal morphology and cover a broad frequency range.
Burst pipelines, such as \gls{cwb}, rely on the coherence of a signal across the detector network rather than on a fixed waveform model~\cite{2005PhRvD..72l2002K,2008CQGra..25k4029K,2016PhRvD..93d2004K,2025PhRvD.111b3054M}.
Therefore, any distortion of the signal that is not consistent with the expected response of the interferometer can degrade the coherence and lead to a loss of sensitivity.
For \gls{cbc} parameter estimation, this distortion is taken into account through frequency dependent calibration errors reconstructed from the median, 16th and 84th quantile of the magnitude and phase errors~\cite{2025arXiv250818081T}.
Additionally, parameter estimation has been performed using the most accurate quantification of the calibration error described in~\cite{2020CQGra..37v5008S}, which showed similar results~\cite{2021PhRvD.103f3016V}.
Historically, burst searches have incorporated calibration errors using highly simplified approximations, typically a global amplitude scaling or a constant time shift~\cite{2016PhRvD..94j2001A,2020PhRvD.101h4002A}.
These approaches approximate the dominant calibration effects for compact binary coalescence signals, which are relatively narrowband, but they fail to capture the frequency‑dependent distortions relevant for broadband \gls{ccsn} waveforms.
Recent \glspl{ccsn} analyses have begun incorporating more realistic calibration uncertainty curves~\cite{2024PhRvD.110d2007S,2025ApJ...985..183A}, but a detailed and systematic study of their impact on burst detection statistics has not yet been performed.

In this work, we introduce a new calibration‑error plugin for \texttt{cWB} that enables the application of physically motivated, frequency‑dependent calibration errors to injected waveforms.
We use this tool to perform the first systematic investigation of how realistic calibration uncertainty affects the detection statistics, detection efficiencies, and astrophysical statements such as the upper limit on the explosion energy of \gls{ccsn} events.
By combining first‑order analytical predictions with extensive injection campaigns using real interferometer data around the explosion time of SN 2023ixf~\cite{2025arXiv250818079T}, we quantify the extent to which calibration uncertainties influence \texttt{cWB} detection statistics and assess whether these effects are astrophysically significant.

\section{Interferometric Systematic Errors and Uncertainty}
\label{sec:calibration_uncertainty}

Because the calibration process is complex and an active research area, the details of the process depend on the specific observatory and the observing run~\cite{2021PTEP.2021eA102A,2026JPhCS3177a2082A,2020CQGra..37v5008S,2021arXiv210700129S}.
This section provides a brief introduction to the calibration process of \gls{gw} interferometers that contains the necessary information to understand the impact of calibration errors on the search pipelines.
In particular, the focus lies on the process followed for the \gls{ligo} observatories during the third and fourth observing runs~\cite{2020CQGra..37v5008S,2021arXiv210700129S}, which is the most representative of the data used in this publication because the event under study, SN 2023ixf, occurred at the start of the fourth observing run.

\begin{figure}
	\centering
	\includegraphics[width=0.8\textwidth]{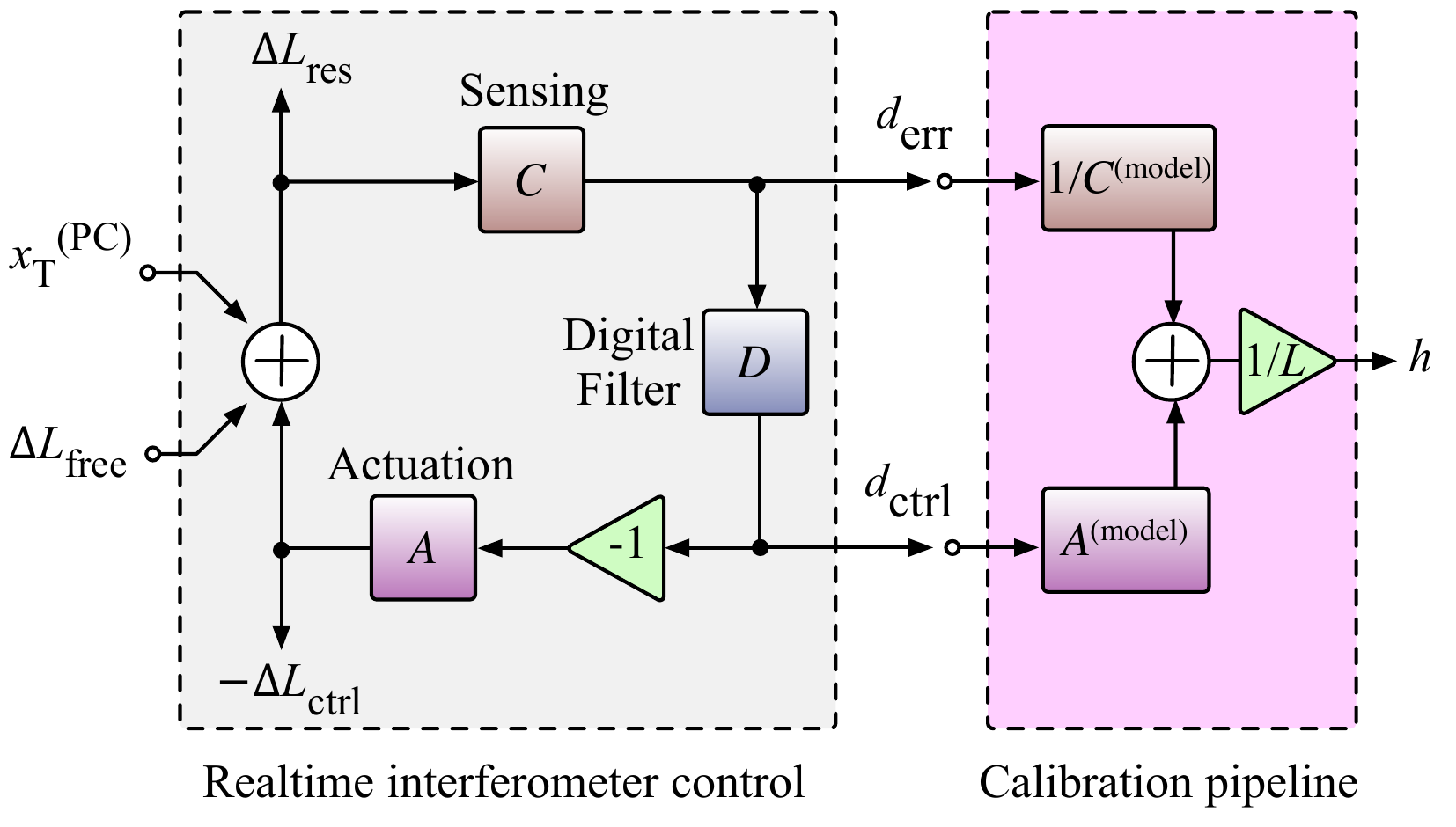}
	\caption{The \gls{ligo} \gls{darm} diagram shows the feedback control loop, shaded in grey, and the calibration process, shaded in pink. The dimensionless strain \emph{h(t)} is obtained by measuring control signals $d_{err}$ and $d_{ctrl}$, using them as inputs to the calibration pipeline~\cite{2017PhRvD..96j2001C}.}
	\label{fig:control_loop}
\end{figure}

Current \gls{gw} detectors use Michelson interferometers to to measure the length difference in the orthogonal arms of the detector, which is then converted to the dimensionless strain \emph{h(t)}.
Therefore, the measurement model can be described conceptually as
\begin{equation}
	\label{eqn:strain}
	h(t) = \frac{L_x - L_y}{L} = \frac{\Delta L_{free}}{L}
	\text{,}
\end{equation}
In practice, the interferometer does not measure the free \gls{darm} length $\Delta L_{free}$ directly but rather a digitised version the electric response to the light intensity variations at the output of the interferometer caused by a stretching of the arm cavities.
This process can be described in the frequency domain as a linear time-invariant system with a sensing transfer function $\tilde{C}(f)$, where $\tilde{\dot}$ represents a frequency domain quantity.
Additionally, the interferometer is kept in resonance by a feedback control loop which applies a control signal to the interferometer to counteract the effect of the \gls{gw} and other noise sources.
Therefore, the measured strain is no longer the direct measurement of the free \gls{darm} length $(\tilde{C} \mathrm{\Delta \tilde{L}_{free}})(f) / L$ but has to be reconstructed using a model for the sensing transfer function $\tilde{C}(f)$ and the actuation transfer function $\tilde{A}(f)$ which describes the response of the interferometer to the control signal.
This process is visualised in figure \ref{fig:control_loop} where the control loop is broken into two distinct blocks.

The grey shaded region, is known as the \gls{darm} feedback control loop and the pink shaded region serves as a representation of the calibration pipeline or ``calibration process'' that yields the final strain measurement \emph{h(t)}.
The \gls{darm} loop contains the physical interferometer, analog electronics, analog-to-digital converters, a network of ``front-end'' computers, and digital-to-analog converters, as described in~\cite{2021SoftX..1300619B}.
The calibration pipeline reconstructs the free \gls{darm} length changes based on two transfer functions, one to invert the sensing function and one to undo the effect of the actuation
\begin{equation}
	\begin{aligned}
		\Delta \tilde{L}_{free}(f) = \Delta \tilde{L}_{res}(f) + \Delta \tilde{L}_{ctrl}(f) =
		\frac{1}{\tilde{C}^{(model)}(f)} \tilde{d}_{err} + \tilde{A}^{(model)} \tilde{d}_{ctrl}
	\end{aligned}
	\text{,}
\end{equation}
where the superscript ``model'' indicates that the transfer functions used in the calibration pipeline are models of the true transfer functions of the interferometer.
By moving to the frequency domain representation of the system where convolutions become multiplications, the reconstructed strain can be written as~\cite{2017PhRvD..95f2003A, 2017PhRvD..96j2001C, 1608.05134, 2018CQGra..35i5015V}
\begin{equation}
	\begin{aligned}
		\label{eqn:h_response}
		\tilde{h}^{(cal)}(f) = \frac{\tilde{R}^{(model)} \tilde{d}_{err}}{L}
	\end{aligned}
	\text{,}
\end{equation}
where
\begin{equation}
	\Tilde{R}^{(model)} = \frac{1 + \Tilde{A}^{(model)}\Tilde{D}\Tilde{C}^{(model)}}{\Tilde{C}^{(model)}}
\end{equation}

If the model for the sensing and actuation transfer functions are not perfect, then $\tilde{R}^{(model)}$ is not identical to the true response of the interferometer $\tilde{R}^{(true)}$ and therefore the reconstructed strain $\tilde{h}^{(cal)}$ is not identical to the true strain $\tilde{h}^{(true)}$.
The calibration model is built for every calibration epoch, which is a period in which the interferometer is in a fixed configuration and the response is expected to be stable.
At the start of each calibration epoch, the transfer functions $\tilde{C}$ and $\tilde{A}$ are fitted using model parameters, e.g. cavity pole frequency, using measured responses and \gls{mcmc} sampling to obtain the posterior distribution of the model parameters.
At the start of each calibration epoch, the model parameters in $\tilde{C}$ and $\tilde{A}$, e.g. the cavity pole frequency, are fitted to measured responses using \gls{mcmc} sampling, which results in a posterior distribution of the model parameters.
The maximum a posteriori parameters of this distribution are then used as the reference model for the calibration of the data during that epoch.
During the operation of the interferometer, calibration lines are injected by a photon calibrator~\cite{2017PhRvD..96j2001C} and / or Newtonian calibrator~\cite{2024CQGra..41w5003A} to track any time-varying changes in the response of the interferometer, such as an amplitude scaling, and to account for them in the calibration model.
Combining the reference model with the time-varying corrections allows for the fitting of \gls{fir} filters to the response model.
These filers are then applied to the interferometer data to produce calibrated strain time series~\cite{2022CQGra..39d5006A,2018CQGra..35i5015V,2025CQGra..42u5016W}.

The resulting calibrated strain has to be interpreted because the calibration model is not guaranteed to be the exact response of the interferometer.
For example, the parameters used in reference model have uncertainties, the model itself might not be a perfect representation of the true response, the time-varying corrections have a measurement uncertainty and the \gls{fir} implementation cannot model the response to infinite precision.
First, weekly measurements of the sensing and actuation transfer functions are collected and compared to the reference model.
Then, a \gls{gpr} is fitted on the ratio between the measured response and the modelled response which yields a distribution over the frequency dependent amplitude and phase errors.
Finally, draws from this distribution are combined with samples from the \gls{mcmc} model for the calibration parameters, time-varying correction uncertainty and photon calibrator uncertainty.
This results in a distribution of the magnitude and phase response of the interferometer.
By dividing them with the reference model, the systematic error and the uncertainty in the response can be determined (see figures~\ref{fig:amp_scale} and~\ref{fig:time_delay}).
To save storage, only a summary of the calibration uncertainty is stored, namely the median, the 16\textsuperscript{th} percentile and the 84\textsuperscript{th} percentile curves of the response distribution.

\section{Coherent WaveBurst}
\label{sec:cwb}

The task of a burst detection pipeline is to classify each stretch of data as noise ($\mathcal{M}_{n}$) or signal plus noise ($\mathcal{M}_{s+n}$).
This work focusses on the traditional approach currently used in the \gls{lvk} collaboration for the all-sky~\cite{2025arXiv250712374L,2025arXiv250712282T}, \gls{ccsn}~\cite{2024PhRvD.110d2007S,2025ApJ...985..183A}, gamma-ray burst~\cite{2022ApJ...928..186A} and fast radio burst~\cite{2023ApJ...955..155A} search.
In particular, the \glsentryfull{cwb} pipeline will be used in the \gls{xp} version but the conclusions should be applicable to any traditional burst detection pipeline due to their similar methodology.
We will not consider machine learning based methods for burst detection which train a classifier directly on the time-domain data.

The major difference between burst pipelines and the ubiquitous matched filtering technique used for the detection of \gls{cbc} signals is that it does not rely on a waveform model.
To be precise, it only assumes that the \gls{gw} signal across the detector network is consistent with general relativity and that the signal energy is concentrated in the time-frequency domain.
The former assumption yields the simple signal model
\begin{equation}
	h_m(t) = F_{m,+} h_{+}(t) + F_{m,\times} h_{\times}(t)
	\text{,}
	\label{eq:signal_model}
\end{equation}
where $h_m(t)$ is the \gls{gw} response in the $m$-th detector, $F_{m,+}$ and $F_{m,\times}$ are the antenna pattern functions for the plus and cross polarizations respectively, and $h_{+}(t)$ and $h_{\times}(t)$ are the plus and cross polarisations of the \gls{gw} signal.
The second assumption is usually satisfied for \gls{gw} signals due to the physical processes that generate them.
For example, the process can have a characteristic frequency such as the orbital frequency, rotational frequency or an oscillation mode (\gls{pns} oscillation or \gls{sasi}).
Alternatively, the process can be transient in nature such as the core bounce in \glspl{ccsn}.

First, the data is transformed into the time-frequency domain because of the aforementioned compact support of \gls{gw} signals.
Different pipelines make different choices for the time-frequency transform.
For example, \texttt{X-Pipeline}~\cite{2010NJPh...12e3034S} and \texttt{PySTAMPAS}~\cite{2021PhRvD.104j2005M} use a short-time Fourier transform while \gls{cwb} 2G uses the \gls{wdm}~\cite{2012JPhCS.363a2032N} transform and \gls{cwb} \gls{xp} uses the wavescan transform~\cite{2022arXiv220101096K}.

Next, the assumption of compact support in the time-frequency domain is used to prune the data and only retain pixels that exceed some statistic that indicates the presence of signal.
In \gls{cwb} \gls{xp}, this done by thresholding and clustering on the cross-power~\cite{2022arXiv220101096K} time-frequency map which is constructed from the individual detector time-frequency maps.

Then a series of detection statistics is computed for each cluster of pixels identified in the previous step.
The most important detection statistic is the (constrained) maximum likelihood ratio~\cite{2005PhRvD..72l2002K, 2008CQGra..25k4029K}
\begin{equation}
	L_{max} = \max_{h_{+}, h_{\times}} \frac{p(\mathbf{d} | h_{+}, h_{\times}, \mathcal{M}_{s+n})}{p(\mathbf{d} | \mathcal{M}_{n})}
	\text{,}
\end{equation}
which can be interpreted as an extension of the Neyman-Pearson likelihood ratio test.
Under the assumption of stationary Gaussian noise, independence of pixels~\footnote{This assumption is usually only approximately satisfied but it allows for the factorisation of the likelihood over pixels. Without it the use of clustering would not be justifiable.}
and uncorrelated noise between detectors, this optimisation problem reduces to finding the solution to the stationarity condition which has an analytical solution
\begin{equation}
	L_{max} = \sum_{i \in C} \boldsymbol{w}[i]^H \mathbf{P}_{sig}[i] \boldsymbol{w}[i]
	\text{,}
	\label{eq:likelihood_ratio}
\end{equation}
where $i$ is the pixel index in the time-frequency map, $C$ is the cluster of pixels associated with the trigger, $\boldsymbol{w}[i]$ is the whitened data in the time-frequency domain and $\mathbf{P}_{sig}[i]$ the projection matrix onto the signal space~\cite{2016PhRvD..93d2004K}.
This projection matrix is where the assumption of coherence across the detector network is enforced because they are dependent on the antenna pattern functions of equation~\ref{eq:signal_model}.

In practice, detector noise is not stationary nor Gaussian due to the presence of glitches.
These glitches will result in triggers with high maximum likelihood ratios.
Therefore additional detection statistics are computed to discriminate between glitches and \gls{gw} signals.
In this study, we will focus on the correlation coefficient ($cc$) and coherent network \gls{snr} ($\rho_{cwb}$) statistics which are derived from the maximum likelihood ratio and the null energy\cite{2008CQGra..25k4029K, 2025PhRvD.111b3054M}.
To obtain these statistics, we first split the maximum likelihood ratio into what we call the coherent and incoherent parts
\begin{equation}
	L_{max} = E_{coh} + E_{i}
	\text{,}
	\label{eq:likelihood_ratio_split}
\end{equation}
where
\begin{align}
	E_{coh} & = \sum_{i \in C } \sum_{m \neq n} L_{mn}[i] \label{eq:coherent_energy} \\
	E_{i}   & = \sum_{i \in C } \sum_{m} L_{mm}[i] \label{eq:incoherent_energy}
	\text{,}
\end{align}
and
\begin{equation}
	L_{mn}[i]  = w_n[i]^H \left( \mathbf{P}_{sig}[i] \right)_{mn} w_n[i]
	\text{,}
	\label{eq:likelihood_ratio_quadratic}
\end{equation}
where $m$ and $n$ are the detector indices.

From equations~\ref{eq:coherent_energy} and~\ref{eq:likelihood_ratio_quadratic}, it can be seen that $E_{coh}$ will be large in the presence of a \gls{gw} signal if the detector responses are similar.
Indeed, if the signal is coherent in both detectors then the terms $m \neq n$ should be relatively large.
This makes $E_{coh}$ a powerful discriminator between signals and glitches for the \gls {ligo} Hanford-Livingston network because of their similar orientations.

In the case of a two-detector network with non-perfectly aligned detectors, the signal model has two degrees of freedom, $h_{+}$ and $h_{\times}$, but the data vector also only has two components, one for each detector.
Therefore, the maximum likelihood solution corresponds to the data vector itself, or equivalently the projection matrix $\mathbf{P}_{sig}[i]$ is the identity matrix.
This is known as the two-detector paradox~\cite{2005PhRvD..72l2002K} and leads to a coherent energy that is zero for any trigger.
Therefore, a heuristic one dimensional signal space is defined by using the data vector as the basis vector so that the same detection statistics can be used for this network configuration.
This corresponds a large coherent energy if the energy is balanced between both detectors and a small coherent energy if the energy is unbalanced between both detectors.

Additionally, the null energy is a useful detection statistic because glitches will leave not be fully reconstructed and leave an unusually high energy in the null stream.
This null energy is defined as the energy after subtracting the reconstructed signal from the data
\begin{equation}
	E_{null} = \sum_{i \in C} (\boldsymbol{w}[i] - \hat{\boldsymbol{w}}[i])^H (\boldsymbol{w}[i] - \hat{\boldsymbol{w}}[i])
	\text{,}
	\label{eq:null_energy_reconstructed}
\end{equation}
where $\hat{\boldsymbol{\xi}}[i] = \hat{\mathbf{P}}_{sig}[i] \boldsymbol{w}[i]$ is the reconstructed signal.
Note that $\hat{\mathbf{P}}_{sig}[i] \neq \mathbf{P}_{sig}[i]$ because the reconstruction uses regulators~\cite{2008CQGra..25k4029K,2016PhRvD..93d2004K} to suppress reconstruction in the insensitive regions of the signal space.

We can rewrite the null energy as
\begin{equation}
	E_{null} = \sum_{i \in C} \boldsymbol{w}[i]^H \hat{\mathbf{P}}_{null}[i] \boldsymbol{w}[i]
	\text{,}
	\label{eq:null_energy_projection}
\end{equation}
where $\hat{\mathbf{P}}_{null}[i] = (\mathbf{I} - \hat{\mathbf{P}}_{sig}[i])^{H} (\mathbf{I} - \hat{\mathbf{P}}_{sig}[i])$ which can no longer be simplified to $\mathbf{I} - \hat{\mathbf{P}}_{sig}[i]$ as in equation~\ref{eq:likelihood_ratio} because $\hat{\mathbf{P}}_{sig}[i]$ is not a real projection matrix and does not satisfy the idempotent property $\hat{\mathbf{P}}_{sig}[i]^{H} \hat{\mathbf{P}}_{sig}[i] = \hat{\mathbf{P}}_{sig}[i]$.

In practice, the coherent energy and null energy are combined into one powerful detection statistic, the correlation coefficient
\begin{equation}
	cc = \frac{E_{coh}}{E_{coh} + E_{n}}
	\text{,}
	\label{eq:network_correlation_coefficient}
\end{equation}
which is close to one for signals and small for glitches.

Finally, the coherent network \gls{snr} is defined as
\begin{equation}
	\rho_{cwb} = \sqrt{
		\frac{E_{coh}}{1 + \overline{\chi}^2 (\max (1, \overline{\chi}^2) - 1)}
	}
	\text{,}
	\label{eq:coherent_network_snr}
\end{equation}
where $\overline{\chi}^2 = E_{n} / N_{DoF}$ is the reduced chi-squared statistic and $N_{DoF}$ is the number of degrees of freedom in the trigger null stream which is proportional to the number of pixels in the trigger.
This serves as an estimate of the network \gls{snr} of the trigger and was used as the ranking statistic in the third observing run of the \gls{lvk} analysis~\cite{2025PhRvD.111b3054M}.

Since the fourth observing run, the aforementioned detection statistics are used as input features for machine leaning classifiers such as XGBoost~\cite{2021PhRvD.104b3014M, 2023PhRvD.107f2002S} or Gaussian Mixture Models~\cite{2020PhRvD.102j4023G, 2022PhRvD.105f3024L, 2024PhRvD.110h3032S} to further discriminate between signals and glitches.
In this paper, we will focus on the estimation of the impact of calibration errors on the detection statistics directly and verify that the impact on the XGBoost classifier is of the same order of magnitude.

\section{Current Methodology}

The presence of calibration errors, as a result of systematic errors and uncertainties in the response function of the interferometer, impacts the likelihood of extracting \glspl{gw} from the data stream as well as the accuracy to which they are reconstructed which leads to uncertainty in the estimation of source and astrophysical parameters.
It is possible however to account and quantify the effect of calibration errors, with respect to simulated \gls{gw} signals from \glspl{ccsn}.
The Burst-Supernova group adopted the methods used for \glspl{gw} emitted from \gls{cbc} which consists of two types of calibration errors.

First, the signal strength is modified by scaling the amplitude of the signal with a constant factor.
Secondly, a time jitter is applied which acts on the phase of the signal.
This method is justified for narrowband signals such as \glspl{cbc} because any transfer function can be approximated in a narrow frequency range by these two effects.

\begin{figure}
	\centering
	\begin{subfigure}{\textwidth}
		\includegraphics[width=\textwidth]{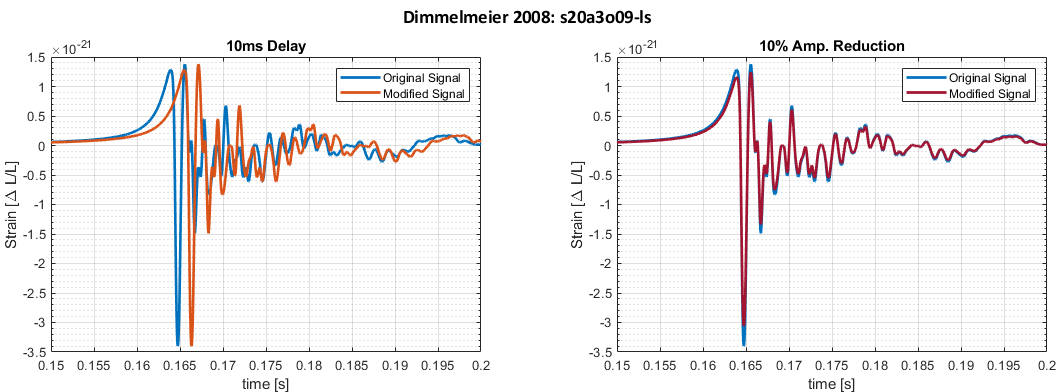}
		\caption{The old methodology to simulate the effect of calibration errors. On the left, the time jitter is shown by applying a 10ms delay to the waveform. On the right, the reduction in signal's strength is achieved by applying a 10\% reduction in the waveform's amplitude.}
		\label{fig:Dim08}
	\end{subfigure}
	\begin{subfigure}{\textwidth}
		\includegraphics[width=\textwidth]{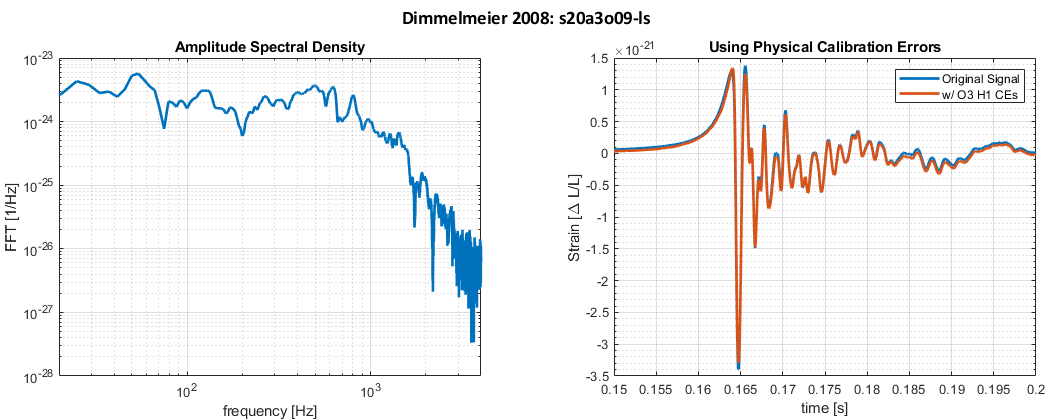}
		\caption{The effect of realistic calibration errors. On the left, the amplitude spectral density of the waveform is plotted to show the broadband nature of \gls{ccsn} Gws. On the right, the original waveform is plotted against, in blue, itself when the physical detector calibration errors are applied, in orange. The calibration errors used is an instantiation picked from a collection at a GPS time during O3 for the H1 detector.}
		\label{fig:Dim08CE}
	\end{subfigure}
	\caption{Example of the effect of different types of calibration errors. The waveform shown is the plus polarization of the waveform s20a3o09, a rapidly rotating \gls{ccsn} \gls{gw} from the Dimmelmeier waveform catalog~\cite{2008PhRvD..78f4056D}.}
\end{figure}

The right side of figure \ref{fig:Dim08} shows the first method: a reduction of the signal's strength by reducing the amplitude component of the \gls{gw} by 10\%.
Conversely, the left side of figure \ref{fig:Dim08} shows an example of time jitter for a delay of \SI{10}{\milli \second}.
Notice that, although the waveforms are modified, the morphology remains the same.
This implies that the effect of the distortion is largely degenerate with signal parameters such as sky location, distance and polarisation.
As a result, only a limited effect is expected on the detectability but the inferred parameters could be biased.

\subsection{Assessing the Suitability}

\begin{figure}
	\centering
	\includegraphics[width=\textwidth]{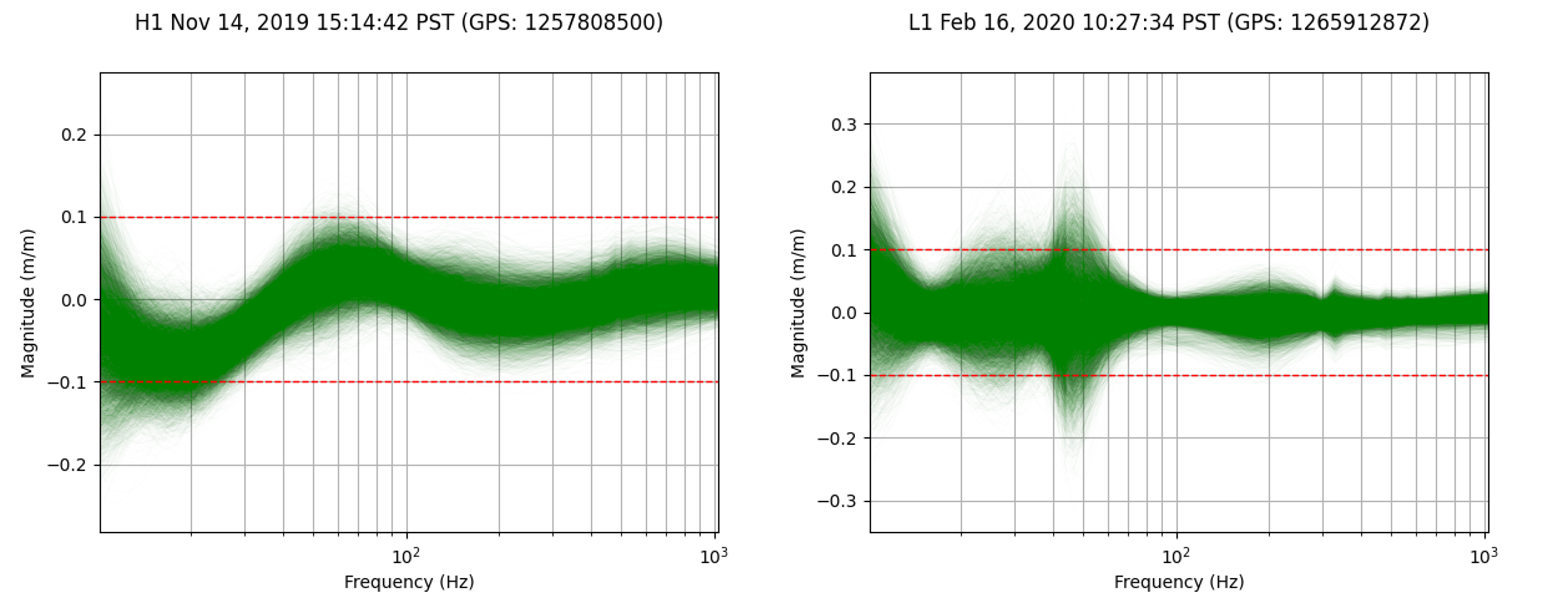}
	\caption{The figure shows the frequency dependence between applying a scale factor of 10\% to the strain amplitude as shown by the red trace and the estimate of the amplitude calibration errors. The green traces show the excursions of the interferometer response model from unity magnitude for a particular GPS during the third observation run. The green traces denoting the frequency dependent errors are taken from a collection a times for when the combined error and uncertainty was the largest during the O3 run.}
	\label{fig:amp_scale}
\end{figure}

Modifying the signal strength or inducing time jitter are methods that have been developed to estimate the impact of calibration errors on \glspl{gw} from \gls{cbc} sources.
These methods are adequate over a narrow frequency range because at a single frequency any linear transformation is fully characterised by an amplitude scaling and phase shift.
This makes this simple method useful for the narrow frequency signatures of \glspl{gw} from \glspl{cbc}.
They however are not suitable when accounting for physical calibration errors across for signals containing energy over a wider range of the detector's frequency band.
Accurately accounting for calibration errors across a broad frequency range is necessary if the impact is to be quantified for broadband \gls{gw} signals, such as those theorize to be emitted from \glspl{ccsn}.
The broadband nature of \glspl{ccsn} is demonstrated by the amplitude spectral density for the Dimmelmeier waveform on the left hand side of \ref{fig:Dim08CE}.
To show that the old methodology is in not well suited, the effect is visualised and compared to a realistic calibration error from the third observing run~\cite{2023ApJS..267...29A}.
First, the method of modifying the signal strength is visualized in figure \ref{fig:amp_scale}.
The red trace shows that a 10\% modification of the signal's strength is obtained by applying a scale factor to the amplitude of the signal achieving either an amplification, using the upper trace, or a reduction, using the lower trace.
The green traces show the excursions of the interferometer response model from unity magnitude, known as the amplitude calibration errors.
These calibration errors are obtained using the methodology described in section~\ref{sec:calibration_uncertainty}.
Note that the green traces denoting the calibration errors are taken from a collection a times for when the combined error and uncertainty was the largest for the Hanford (H1) and Livingston (L1) detectors.
In terms of the induced amplitude error as a function of frequency, the 10\% scale factor performs well in terms of magnitude at some select frequencies but fails to capture the frequency dependence present in the amplitude calibration errors.

\begin{figure}
	\centering
	\includegraphics[width=\textwidth]{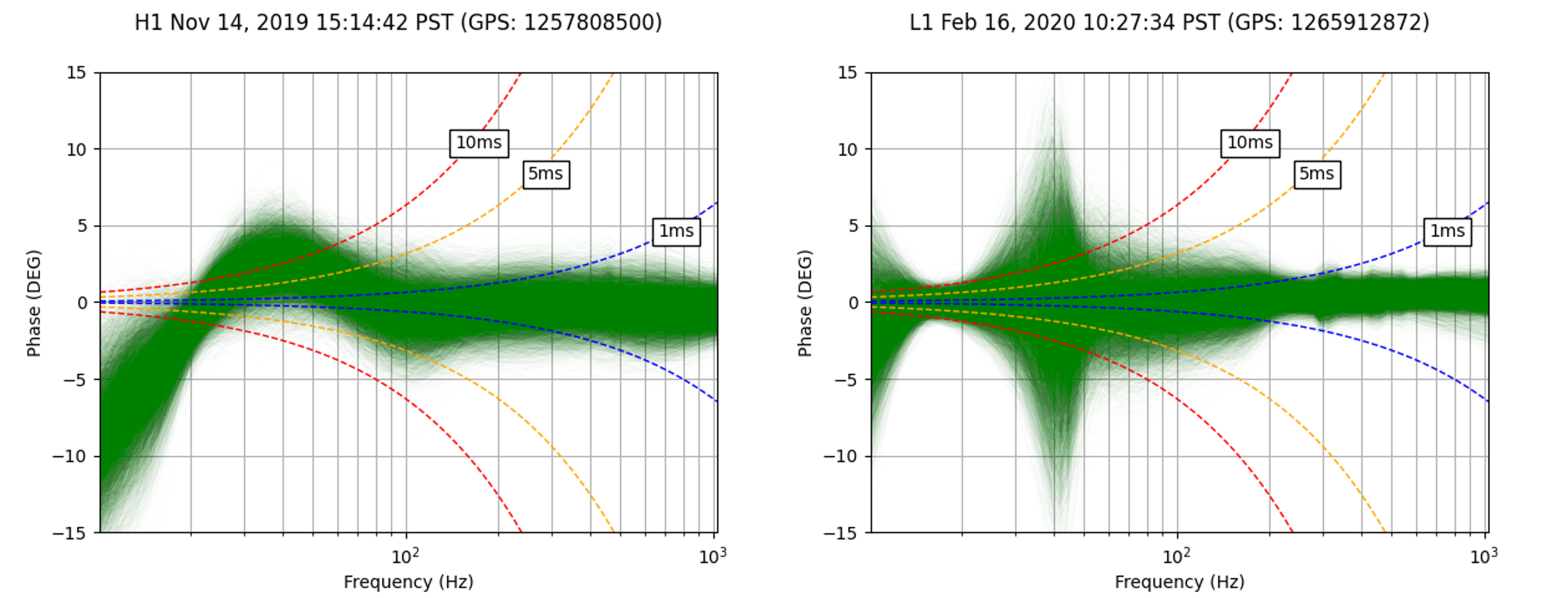}
	\caption{The figure shows the frequency dependence of applying a time delay, using the lower traces, or an advance, using the upper traces, of either 1ms, blue trace, 5ms, yellow trace, or 10ms, red trace. Compared to the green traces which show the excursions of the interferometer response model from zero phase for a particular GPS during the third observation run, known as the phase calibration errors. The green traces denoting the frequency dependent errors are taken from a collection a times for when the combined error and uncertainty was the largest during the O3 run.}
	\label{fig:time_delay}
\end{figure}

Secondly, the method of time jitter is visualized in figure \ref{fig:time_delay}, blue, yellow and red traces denote a 1ms, 5ms, and 10ms time jitter respectively; achieving a delay using the negative sloping curves and an advance using the positive slopping curves.
Notice that a time delay or advance of a signal induces a phase shift that is linear as a function of frequency and scales with respect to the magnitude of the jitter applied.
The green traces show the excursions of the interferometer response model from zero phase, known as the phase calibration errors.
The frequency dependence of the phase calibration errors does not follow the same linear evolution over frequency.
Figure \ref{fig:Dim08CE}, shows the effect of calibration errors on the Dimmelmeier waveform using an instantiation of the H1 amplitude and phase errors from the curves shown in figure \ref{fig:amp_scale} and figure \ref{fig:time_delay} derived at a GPS time during O3.
Comparing figure~\ref{fig:Dim08} and figure~\ref{fig:Dim08CE}, it can be seen that physical calibration errors can induce a distortion in the morphology of the waveform which is not captured by the old methodology.

\subsection{Impact on previously published results}

Calibration errors have been taken into account in previous \gls{ccsn} search papers through a very conservative estimation of the impact on detection efficiencies~\cite{2016PhRvD..94j2001A, 2020PhRvD.101h4002A}.
The rationale behind this conservative estimation~\cite{2009PhRvD..80j2002A} is that in the worst case scenario, as the amplitude calibration is constant and equal to the maximum amplitude calibration error across the frequency band.
Assuming that the amplitude calibration errors are independent in each detector, the variance of the network \gls{snr} can be estimated.
As a reduction in network \gls{snr} is similar to an increase in distance, calibration errors can be accounted for by reducing the distance limits at a fixed detection efficiency by a factor corresponding to the 90\% confidence interval of the \gls{snr} variance.
For example, in the search for \glspl{ccsn} during the first and second observing run amplitude calibration errors were estimated to be 10\% for both H1 and L1 at the times of the five \glspl{ccsn} studied.
This resulted in a decision to decrease distance limits at a fixed detection efficiency by 9.1\%.
As stated in~\cite{2020PhRvD.101h4002A}, the dominant effect on detection efficiencies was understood to be from the uncertainties in the strain amplitude calibration, as in~\cite{2016PhRvD..94j2001A}.
Therefore, the effect of phase calibration errors was ignored in these previous search papers even though they can induce a distortion in the morphology of the waveform which can impact the detection statistics of burst pipelines.

The O3 search paper~\cite{2024PhRvD.110d2007S} already used the methodology presented in this paper to simulate the effect of calibration uncertainty.
This led to a more accurate estimation of the impact of calibration errors on detection efficiencies and explosion energy upper limits.
It was observed that the impact of calibration errors on the detection efficiencies of core-collapse supernovae was negligible compared to the other uncertainties in the detection efficiencies.
In this paper we will describe this methodology in detail and do a more systematic investigation of the impact of calibration errors on detection statistics, detection efficiencies and explosion energy upper limits.
The SN2023ixf search paper~\cite{2025ApJ...985..183A} already includes a statement about the negligible impact of calibration errors on detection efficiencies which is based on the results of the analysis presented in this paper.

\section{Methodology}

The study aims to achieve two primary objectives.
First, to investigate the impact of realistic frequency-dependent calibration errors on detection statistics for broadband burst signals.
Secondly, to examine how calibration uncertainties affect scientific statements for \gls{ccsn} searches, including detection efficiencies and explosion energy upper limits.
While this work focusses on \gls{cwb}, the methodology can easily be extended to other burst pipelines in the future.

\subsection{Simulation setup}
\label{sec:simulation_setup}

The simulation setup involves running \gls{cwb} in simulation mode, where numerically simulated \gls{ccsn} waveforms are injected into the strain data.
This study employs real interferometric data within a simulation framework identical to that used in the SN 2023ixf search~\cite{2025ApJ...985..183A}, i.e. it uses data from the Hanford and Livingston detectors from 13 May 2023 19:49:35 UTC to 18 May 2023 19:49:35 UTC when both detectors are in an observing state.
Since the on source window occurs during the engineering run before the start of the fourth observing run, the detectors were not yet in their nominal state.
As a result, the sensitivity of the detectors was at the lower end of the expected sensitivity for the fourth observing run~\cite{2025PhRvD.111f2002C}.
Additionally, the calibration uncertainties were larger than the typical calibration uncertainties during the fourth observing run.
Within the frequency band used in the search (\SI{32}{\hertz} to \SI{2048}{\hertz}), the maximum amplitude calibration error was around 6\% for the L1 detector and around 14\% for the H1 detector (see figure~\ref{fig:summary}), compared to the typical amplitude calibration errors of around 3\% for the L1 detector and around 5\% for the H1 detector during the fourth observing run.
For the phase calibration errors, the maximum phase error was around \ang{6} for the L1 detector and around \ang{8} for the H1 detector (see figure~\ref{fig:summary}), compared to the typical phase calibration errors of around \ang{3} for the L1 detector and around \ang{4} for the H1 detector during the fourth observing run.

The calibration error plugin, detailed in section~\ref{sec:cwb_plugin}, is utilized to interrupt the normal program flow just before waveforms are added to the noise.
After applying the calibration errors, the normal flow of \gls{cwb} resumes and event triggers are generated.
The simulation is executed twice: once with calibration errors and once without.
Generated triggers from both simulations are then matched based on their GPS times, a process feasible due to the simulation's reproducibility with a fixed seed for the random number generators.
The impact on detection statistics for the matched triggers is investigated, with results presented in section~\ref{sec:impact_detection_statistics}.
Additionally, the impact on detection efficiencies and explosion energy upper limits, following \gls{xgboost} classification, is examined, with results detailed in section~\ref{sec:impact_detection_efficiencies}.
Note that the same \gls{xgboost} model is used for classification in both simulations.
Ideally, this model should be trained on data that has also been affected by calibration errors to ensure that the model is trained on data that is representative of the data it will be applied to but given the small impact of calibration errors on detection statistics in the results of section~\ref{sec:impact_detection_statistics}, we do not expect a significant impact on the classification performance.

\subsection{Calibration error plugin}
\label{sec:cwb_plugin}

A plugin was developed to address and yield a more accurate estimate of the impact of detector calibration errors on reconstruction parameters for GW signals.
The plugin exists as a tool to extend the capabilities of the \gls{cwb} pipeline.
It interrupts the normal flow of the program just before the waveforms are added to the noise and applies a user specified calibration error to the waveforms and-or signals.
The calibration error can be specified in several different ways to allow for a range of different use cases.

The first three options are a frequency independent scale factor, phase shift or time jitter, which are similar to the methods used in previous search papers.
These can be combined in general calibration error of the form
\begin{align}
	\delta A(f)    & = \delta A                        \\
	\delta \phi(f) & = \delta \phi  + 2 \pi f \delta t
	\text{,}
\end{align}
where $\delta A$ is the amplitude scale factor, $\delta \phi$ is the phase shift and $\delta t$ is the time jitter.
The parameters of the calibration error are determined for each injection depending on the configuration chose by the user.
First, $(\epsilon_A, \epsilon_\phi, \epsilon_t)$ determine the magnitude of the calibration error for each of the three components.
Then, the user can chose between a fixed error $\epsilon$ for each injection or a random error drawn from a uniform or Gaussian distribution with range $(-\epsilon, \epsilon)$ or standard deviation $\epsilon$ respectively.

\begin{figure}
	\centering
	\begin{subfigure}{.49\textwidth}
		\includegraphics[width=\textwidth]{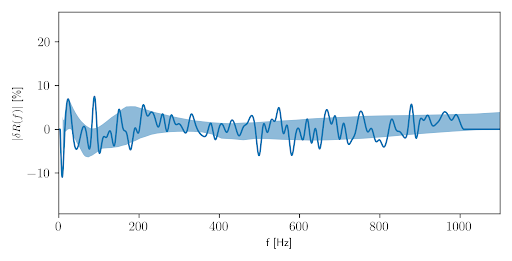}
	\end{subfigure}
	\begin{subfigure}{.49\textwidth}
		\includegraphics[width=\textwidth]{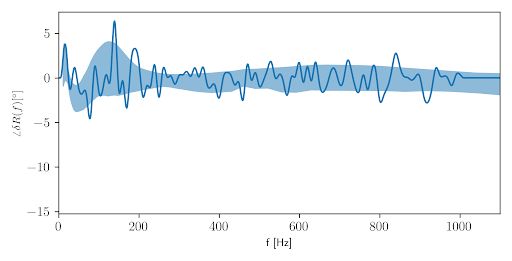}
	\end{subfigure}
	\caption{An example amplitude (left) and phase (right) error generated using the summary files for the H1 detector at GPS time 1368354250. The blue curve shows the generated calibration error and the shaded region shows the 1-$\sigma$ band calibration uncertainty. A frequency resolution of \SI{10}{\hertz} is used over the frequency range between \SI{10}{\hertz} and \SI{1}{\kilo \hertz}.}
	\label{fig:summary}
\end{figure}

The latter three options allow for a more realistic frequency dependent calibration error to be applied to the waveforms.
First, the \texttt{file} option reads in a user specified file containing the frequency dependent amplitude and phase errors.
Secondly, the \texttt{hdf5} option reads in a file containing a collection of realisations of the calibration error curves generated by the procedure outlined in section~\ref{sec:calibration_uncertainty}.
For each injection a random realisation is drawn from the collection and applied to the waveform.
Lastly. the \texttt{summary} option requires the path to a directory containing the summary files with the 16\textsuperscript{th}, 50\textsuperscript{th} and 84\textsuperscript{th} percentiles of the curves used by the \texttt{hdf5} option.
This is useful because the hdf5 files are an intermediate product of the calibration uncertainty estimation procedure and are not always available for all GPS times of interest.
For each injection, the closest GPS time for which the summary files are available is chosen and the confidence intervals are used to model the calibration error as a Gaussian distribution at each frequency.
In this paper, the calibration uncertainty is generated between \SI{10}{\hertz} and \SI{3000}{\hertz} with a frequency resolution of \SI{10}{\hertz}.
For each of the chosen frequencies, the amplitude and phase errors are then determined as follows
\begin{align}
	\delta A(f_i)    & \sim \mathcal{N}(\mu_A(f_i), \sigma_A(f_i))       \\
	\delta \phi(f_i) & \sim \mathcal{N}(\mu_\phi(f_i), \sigma_\phi(f_i))
	\text{,}
\end{align}
where the 50\textsuperscript{th} percentile of the amplitude and phase errors at closest frequency to $f_i$ are used as the mean and half of the difference between the 84\textsuperscript{th} and 16\textsuperscript{th} percentiles are used as the standard deviation.
Figure~\ref{fig:summary} shows an example of the generated calibration error curves using the summary files for the H1 detector during the on-source window of the SN2023ixf search.
For all three option discussed above, the final calibration error is obtained by interpolating the generated curves using a cubic b-spline.

Note that the different options do not only allow for different types of calibration errors but also ensure that different questions can be answered.
For example, using the \texttt{file} option allows for the impact of a specific calibration \emph{error} to be investigated while using the \texttt{hdf5} or \texttt{summary} options test the impact of the \emph{uncertainty} in the calibration.
By drawing different curves for every injection, the simulation marginalises over the calibration uncertainty.

To apply the frequency dependent calibration error to the waveforms, two frequency domain methods are implemented.
The first method is the well-known \gls{ola} which is exact for a transfer function that corresponds to a finite impulse response.
In general, the transfer function is not guaranteed to be a finite impulse response and the \gls{ola} method can introduce circular convolution artefacts.
To mitigate this issue, \gls{wola} is implemented which applies an analysis and synthesis window in the transformation.
This removes the equivalence with finite impulse response filters but suppresses circular convolution artefacts.

\section{Impact on detection statistics}
\label{sec:impact_detection_statistics}

The goal of this section is twofold: first, to calculate the effect of calibration errors on the detection statistics of \gls{cwb} (see section~\ref{sec:cwb}) to first order in the calibration errors. To achieve this goal, the effect of calibration errors on the wavelet domain data is modelled in sections~\ref{sec:calibration_model} and first order calculations are presented in section~\ref{sec:first_order_impact}. Secondly, these predictions are compared to the detection statistics in the simulations described in section~\ref{sec:simulation_setup}, with results presented in section~\ref{sec:simulation_detection_statistics}.

\subsection{Calibration model}
\label{sec:calibration_model}

To model the effect of calibration errors in the wavelet domain, we start from the description of the calibration errors in the frequency domain (see section~\ref{sec:calibration_uncertainty})
\begin{equation}
	\tilde{C}_n(f) = \left( 1 + \delta A_n(f) \right) e^{i \delta \phi_n(f)}
	\text{,}
	\label{eq:calibration_model}
\end{equation}
where $\delta A_n(f)$ and $\delta \phi_n(f)$ are the frequency-dependent amplitude and phase errors respectively for the $n$-th detector.
For small amplitude and phase errors, this can be approximated as
\begin{align}
	\tilde{C}_n(f) & \approx 1 + \delta A_n(f) + i \delta \phi_n(f) \\
	               & = 1 + \delta \tilde{C}_n(f)
	\text{.}
	\label{eq:calibration_model_approx}
\end{align}
The whitened gravitational wave response in the wavelet domain in the presence of calibration errors can then be written as
\begin{align}
	\xi^{CE}_n[i] & = \left< \phi_i(t), \xi^{CE}_n(t) \right> \nonumber                                                          \\
	              & = \int_{-\infty}^{\infty} df \tilde{\phi}_i^*(f) \tilde{\xi}^{CE}_n(f) \nonumber                             \\
	              & = \int_{-\infty}^{\infty} df \tilde{\phi}_i^*(f) (1 + \delta \tilde{C}_n(f)) \tilde{\xi}_n(f) \nonumber      \\
	              & = \xi_n[i] + \int_{-\infty}^{\infty} df \tilde{\phi}_i^*(f) \delta \tilde{C}_n(f) \tilde{\xi}_n(f) \nonumber \\
	              & = \xi_n[i] + \delta \xi_n[i]
	\text{,}
	\label{eq:calibrated_response}
\end{align}
where $\phi_i$ is the wavelet basis function for pixel $i$ and $\xi_n$ is the perfectly calibrated whitened response.
The last term in equation~\ref{eq:calibrated_response} represents the error in the whitened response due to calibration errors.
For complex wavelets, this term can be easily interpreted in terms of the amplitude and phase errors at the central frequency of the wavelet
\begin{align}
	\delta \xi_n[i] & = \int_{-\infty}^{\infty} df \tilde{\phi}_i^*(f) \delta \tilde{C}_n(f) \tilde{\xi}_n(f) \nonumber         \\
	                & \approx \delta \tilde{C}_n(f_i) \int_{-\infty}^{\infty} df \tilde{\phi}_i^*(f) \tilde{\xi}_n(f) \nonumber \\
	                & \approx \delta \tilde{C}_n(f_i) \xi_n[i]
	\text{,}
	\label{eq:calibrated_response_approx}
\end{align}
where it is assumed that the calibration errors vary slowly over the bandwidth of the wavelet centered at frequency $f_i$ and that the wavelet is an analytic signal, i.e. it has no negative frequency content~\footnote{
	If this assumption is not satisfied then it is not possible to move $\delta \tilde{C}_n$ outside of the integral and an approach similar to the real wavelets would have to be used.
}.
Therefore, the error term is proportional to the whitened response itself and the calibration error at the central frequency of the wavelet.
For real wavelets, the error term is no longer proportional to the wavelet coefficient itself because of the negative frequency contribution in the integral
\begin{align}
	\delta \xi_n[i] & = \int_{-\infty}^{\infty} df \tilde{\phi}_i^*(f) \delta \tilde{C}_n(f) \tilde{\xi}_n(f) \nonumber                                                                                                                                                                       \\
	                & = 2 \Re \left[ \int_{0}^{\infty} df \tilde{\phi}_i^*(f) \delta \tilde{C}_n(f) \tilde{\xi}_n(f) \right] \nonumber                                                                                                                                                        \\
	                & \approx 2 \Re \left[ \delta \tilde{C}_n(f_i) \int_{0}^{\infty} df \tilde{\phi}_i^*(f) \tilde{\xi}_n(f) \right] \nonumber                                                                                                                                                \\
	                & \approx \Re \left[ \delta \tilde{C}_n(f_i) \right] \Re \left[ 2 \int_{0}^{\infty} df \tilde{\phi}_i^*(f) \tilde{\xi}_n(f) \right] - \Im \left[ \delta \tilde{C}_n(f_i) \right] \Im \left[ 2 \int_{0}^{\infty} df \tilde{\phi}_i^*(f) \tilde{\xi}_n(f) \right] \nonumber \\
	                & \approx \delta A_n (f) \xi_n[i] - \delta \phi_n(f) \hat{\xi}_n[i]
	\text{,}
	\label{eq:calibrated_response_real}
\end{align}
where $\hat{\xi}_n[i]$ is the wavelet coefficient of the Hilbert transformed wavelet basis function.
This corresponds to a $90^\circ$ phase shift of the original wavelet, also known as the quadrature wavelet.
For the Wilson-Daubechies-Meyer (WDM) wavelet basis used in \texttt{cWB}, the quadrature wavelet coefficients can be computed jointly with the original wavelet coefficients during the wavelet transform.
In fact, the 2G version of \texttt{cWB} uses the quadrature wavelet coefficients to apply polarisation constraints during the reconstruction of the signal~\cite{2016PhRvD..93d2004K}.
In the remainder of this text, we will use the generic form of equation~\ref{eq:calibrated_response} which is first order in the calibration error and the signal strength.

\subsection{First order impact on detection statistics}
\label{sec:first_order_impact}

The impact of calibration errors on the detection statistics for a trigger can be calculated using the formulae in section~\ref{sec:cwb} combined with the calibration model in equation~\ref{eq:calibrated_response}.
First, the null energy in equation~\ref{eq:null_energy_projection} can be expanded to
\begin{align}
	E_{n}^{CE} & = \sum_{i \in C } \left( \boldsymbol{w}^{CE}[i] \right)^H \hat{\mathbf{P}}_{null}[i] \boldsymbol{w}^{CE}[i] \nonumber                                                                \\
	           & = \sum_{i \in C } \left( \boldsymbol{w}[i] + \delta \boldsymbol{\xi}[i] \right)^H \hat{\mathbf{P}}_{null}[i] \left( \boldsymbol{w}[i] + \delta \boldsymbol{\xi}[i] \right) \nonumber \\
	           & = E_{n} + 2 \Re \left[ \sum_{i \in C } \boldsymbol{w}[i]^H \hat{\mathbf{P}}_{null}[i] \delta \boldsymbol{\xi}[i] \right] + \mathcal{O}(\delta \boldsymbol{\xi}^2)
	\text{.}
	\label{eq:null_energy_ce}
\end{align}
Similarly, the maximum likelihood ratio in equation~\ref{eq:likelihood_ratio} becomes
\begin{align}
	L_{max}^{CE} & = \sum_{i \in C} \left( \boldsymbol{w}[i] + \delta \boldsymbol{\xi}[i] \right)^H \mathbf{P}_{sig}[i] \left( \boldsymbol{w}[i] + \delta \boldsymbol{\xi}[i] \right) \nonumber \\
	             & = L_{max} + 2 \Re \left[ \sum_{i \in C } \boldsymbol{w}[i]^H \mathbf{P}_{sig}[i] \delta \boldsymbol{\xi}[i] \right] + \mathcal{O}(\delta \boldsymbol{\xi}^2)
	\text{.}
	\label{eq:likelihood_ratio_ce}
\end{align}
Using the definition of the projection matrix~\cite{2016PhRvD..93d2004K} as the projection onto the dominant polarisation subspace
\begin{equation}
	\mathbf{P}_{sig}[i] =
	\begin{bmatrix}
		\mathbf{e}_+[i] & \mathbf{e}_\times[i]
	\end{bmatrix}
	\begin{bmatrix}
		\mathbf{e}_+[i]
		\mathbf{e}_\times[i]
	\end{bmatrix}^H
	\text{,}
\end{equation}
where $\mathbf{e}_+[i]$ and $\mathbf{e}_\times[i]$ are the normalised dominant polarisation vectors corresponding to pixel $i$, the incoherent energy can be written as
\begin{align}
	E_{i}^{CE} & = \sum_{i \in C } \sum_{m} \left( w_m^{CE}[i] \right)^H (e_{m,+}[i])^2 w_m^{CE}[i] + \left( w_m^{CE}[i] \right)^H (e_{m,\times}[i])^2 w_m^{CE}[i] \nonumber                                                                            \\
	           & = \sum_{i \in C} \left( \boldsymbol{w}[i] + \delta \boldsymbol{\xi}[i] \right)^H \mathbf{D}[i] \left( \boldsymbol{w}[i] + \delta \boldsymbol{\xi}[i] \right) \nonumber                                                                 \\
	           & = \underbrace{\sum_{i \in C} \boldsymbol{w}[i]^H \mathbf{D}[i] \boldsymbol{w}[i]}_{E_{i}} + 2 \Re \left[ \sum_{i \in C } \boldsymbol{w}[i]^H \mathbf{D}[i] \delta \boldsymbol{\xi}[i] \right] + \mathcal{O}(\delta \boldsymbol{\xi}^2)
	\text{,}
	\label{eq:incoherent_energy_ce}
\end{align}
where $e_{m,+}[i]$ and $e_{m,\times}$ are the elements of the normalised dominant polarisation vectors corresponding to $m$-th detector and $\mathbf{D}[i]$ is a diagonal matrix with elements $(\mathbf{D}[i])_{mm} = (e_{m,+}[i])^2 + (e_{m,\times}[i])^2$.
Note that this formula is not valid in the case of two non co-aligned detectors because the projection matrix is defined differently in that case.
In fact, the projection matrix is data dependent and therefore also affected by calibration errors.
It turns out that the formulation of equation~\ref{eq:incoherent_energy_ce} is still valid with an adjusted definition $\mathbf{D}[i]$ which is derived in appendix~\ref{sec:incoherent_energy_derivation}.

Combining this expression with equation~\ref{eq:likelihood_ratio_split} immediately gives expression for the coherent energy
\begin{align}
	E_{coh}^{CE} & = L_{max}^{CE} - E_{i}^{CE} \nonumber                                                                                                                                                       \\
	             & = E_{coh} + 2 \Re \left[ \sum_{i \in C } \boldsymbol{w}[i]^H \left( \mathbf{P}_{sig}[i] - \mathbf{D}[i] \right) \delta \boldsymbol{\xi}[i] \right] + \mathcal{O}(\delta \boldsymbol{\xi}^2)
	\text{.}
	\label{eq:coherent_energy_ce}
\end{align}

Combining equations~\ref{eq:null_energy_ce} and \ref{eq:coherent_energy_ce} with equation~\ref{eq:network_correlation_coefficient} gives the network correlation coefficient in the presence of calibration errors
\begin{align}
	\delta cc^{CE} & \approx \frac{\partial cc}{\partial E_{coh}} \delta E_{coh}^{CE} + \frac{\partial cc}{\partial E_{n}} \delta E_{n}^{CE} \nonumber                                                                                                                                                                                       \\
	               & = \frac{E_{n}}{( E_{coh} + E_{n} )^2} \delta E_{coh}^{CE} - \frac{E_{coh}}{( E_{coh} + E_{n} )^2} \delta E_{n}^{CE} \nonumber + \mathcal{O}(\delta \boldsymbol{\xi}^2)                                                                                                                                                  \\
	               & = \frac{2}{( E_{coh} + E_{n} )^2} \Re \left[ E_{n} \sum_{i \in C } \boldsymbol{w}[i]^H \left( \mathbf{P}_{sig}[i] - \mathbf{D}[i] \right) \delta \boldsymbol{\xi}[i] - E_{coh} \sum_{i \in C } \boldsymbol{w}[i] \hat{\mathbf{P}}_{null}[i] \delta \boldsymbol{\xi}[i] \right] + \mathcal{O}(\delta \boldsymbol{\xi}^2)
	\text{.}
	\label{eq:network_correlation_coefficient_ce}
\end{align}
In the low \gls{snr} regime, all terms in equation~\ref{eq:network_correlation_coefficient_ce} are noise dominated except for the error term $\delta \boldsymbol{\xi}$.
Therefore, the effect of calibration errors will scale with $\boldsymbol{\xi}$ in this regime.
For medium \gls{snr} signals, $E_{coh}$ and $\boldsymbol{w}$ are signal dominated while $E_{n}$ and $\boldsymbol{w}^H \hat{\mathbf{P}}_{null}$ are noise dominated because only the signal component that is suppressed by the regulators will contribute.
Hence, the effect on the correlation coefficient will scale with $\boldsymbol{\xi}^{-1}$ in this regime.
Finally, all terms in equation~\ref{eq:network_correlation_coefficient_ce} are signal dominated for high \gls{snr} signals and the effect of calibration errors will reach an asymptotic value.
Note that because of the limited leakage of signal power to the null energy the asymptotic value is expected to be small and the effect of calibration errors is suppressed for high \gls{snr} signals.
In the two detector case, the projection matrix itself is data dependent but for high \gls{snr} signals it is invariant to a scaling of the signal and the statement above still holds.

Finally, the coherent network \gls{snr} in equation~\ref{eq:coherent_network_snr} has to be studied in two regimes separately: $\overline{\chi}^2 < 1$ and $\overline{\chi}^2 \geq 1$.
If $\overline{\chi}^2 < 1$, we have
\begin{align}
	\rho_{cwb}^{CE} & = \sqrt{E_{coh}^{CE}} \nonumber                                                                                                                                                                                         \\
	                & \approx \sqrt{E_{coh}} + \frac{1}{2 \sqrt{E_{coh}}} \delta E_{coh}^{CE} + \mathcal{O}(\delta \boldsymbol{\xi}^2) \nonumber                                                                                              \\
	                & \approx \rho_{cwb} + \frac{1}{\rho_{cwb}} \Re \left[ \sum_{i \in C } \boldsymbol{w}[i]^H \left( \mathbf{P}_{sig}[i] - \mathbf{D}[i] \right) \delta \boldsymbol{\xi}[i] \right] + \mathcal{O}(\delta \boldsymbol{\xi}^2)
	\text{.}
	\label{eq:coherent_network_snr_ce_low_chi2}
\end{align}
Given that all terms in equation~\ref{eq:coherent_network_snr_ce_low_chi2}, except $\delta \boldsymbol{\xi}[i]$, change from noise dominated to signal dominated as soon as the signal strength becomes comparable to the noise, the effect of calibration errors scales with $\boldsymbol{\xi}$ in this regime.
Therefore, the relative error $\delta \rho_{cwb} / \rho_{cwb}$ increases with \gls{snr} for low \gls{snr} signals and reaches an asymptotic value for medium \gls{snr} signals.
For $\overline{\chi}^2 \geq 1$, we have
\begin{align}
	\delta \rho_{cwb} & \approx \frac{\partial \rho_{cwb}}{\partial E_{coh}} \delta E_{coh}^{CE} + \frac{\partial \rho_{cwb}}{\partial E_{n}} \delta E_{n}^{CE} \nonumber                                                                                                                                                                                                                                                                                         \\
	                  & = \frac{1}{2 \rho_{cwb} (1 + \overline{\chi}^2 (\overline{\chi}^2 - 1))} \delta E_{coh}^{CE} - \frac{E_{coh} (2 \overline{\chi}^2 - 1)}{2 \rho_{cwb} N_{DoF} (1 + \overline{\chi}^2 (\overline{\chi}^2 - 1))^2} \delta E_{n}^{CE} + \mathcal{O}(\delta \boldsymbol{\xi}^2) \nonumber                                                                                                                                                      \\
	                  & = \sum_{i \in C } \Re \left[ \frac{\boldsymbol{w}[i]^H \left( \mathbf{P}_{sig}[i] - \mathbf{D}[i] \right) \delta \boldsymbol{\xi}[i]}{\sqrt{E_{coh}} (1 + \overline{\chi}^2 (\overline{\chi}^2 - 1))} - \frac{\sqrt{E_{coh}} (2 \overline{\chi}^2 - 1) \mathbf{w}[i]^H \hat{\mathbf{P}}_{null}[i] \delta \boldsymbol{\xi}[i]}{N_{DoF} (1 + \overline{\chi}^2 (\overline{\chi}^2 - 1))^2} \right] + \mathcal{O}(\delta \boldsymbol{\xi}^2)
	\text{.}
	\label{eq:coherent_network_snr_ce_high_chi2}
\end{align}
Following a similar argument as for the $\overline{\chi}^2 < 1$ case, $\delta \rho_{cwb} / \rho_{cwb}$ increases with \gls{snr} for low and medium \gls{snr} signals and reaches an asymptotic value when the null energy becomes signal dominated.

\subsection{Simulation detection statistics}
\label{sec:simulation_detection_statistics}

This section presents the results of the simulations described in section~\ref{sec:simulation_setup} and compares them to the predictions from the first order calculations in section~\ref{sec:first_order_impact}.
Two sets of simulations are performed: one with numerically simulated \gls{ccsn} waveforms and one with sine-Gaussian waveforms, which have a more compact time-frequency representation.

\begin{figure}
	\centering
	\begin{subfigure}{0.5\textwidth}
		\centering
		\includegraphics[width=\textwidth]{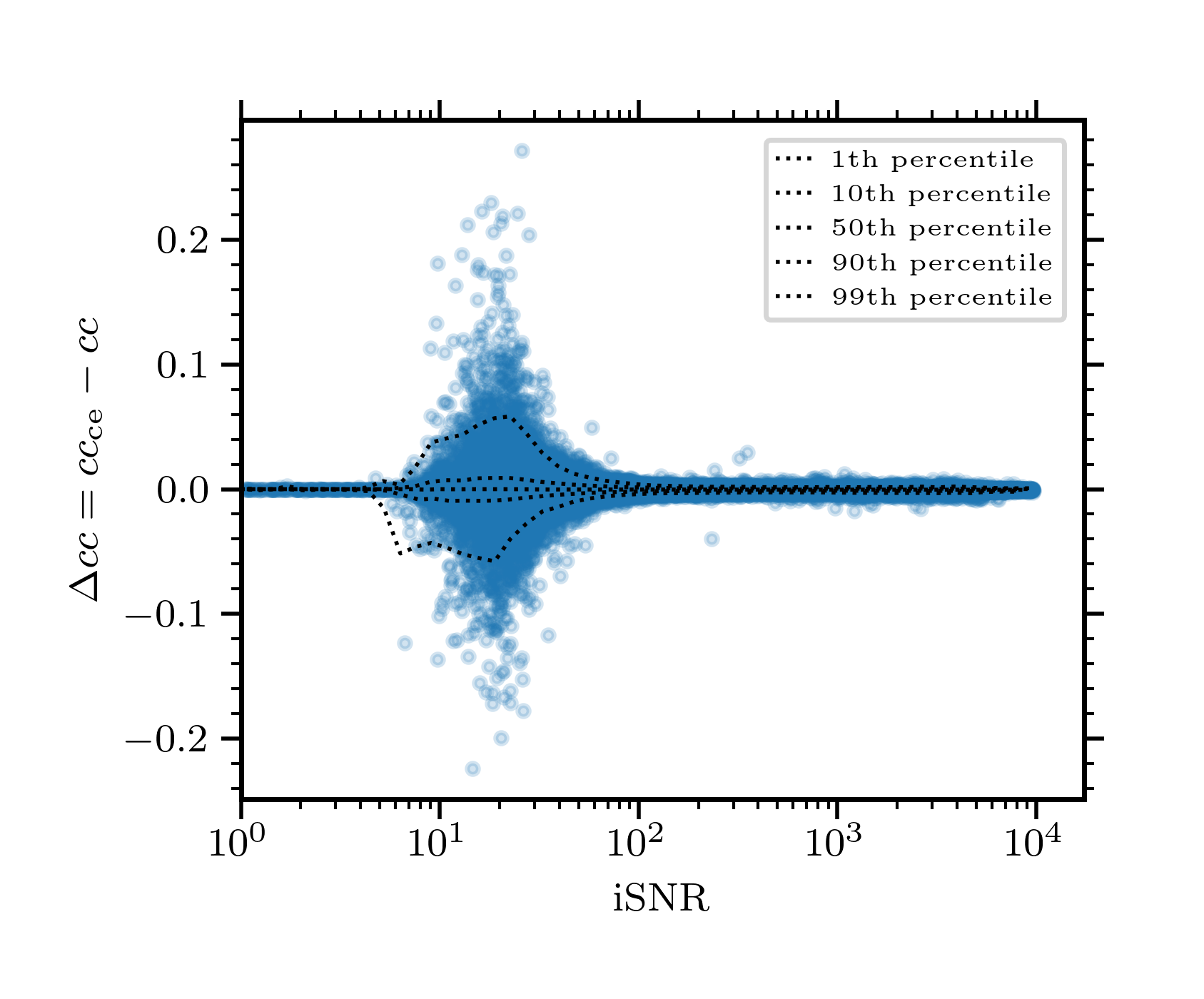}
		\caption{Absolute change in correlation coefficient $cc$}
		\label{fig:cc_vs_snr_sub}
	\end{subfigure}%
	\begin{subfigure}{0.5\textwidth}
		\centering
		\includegraphics[width=\textwidth]{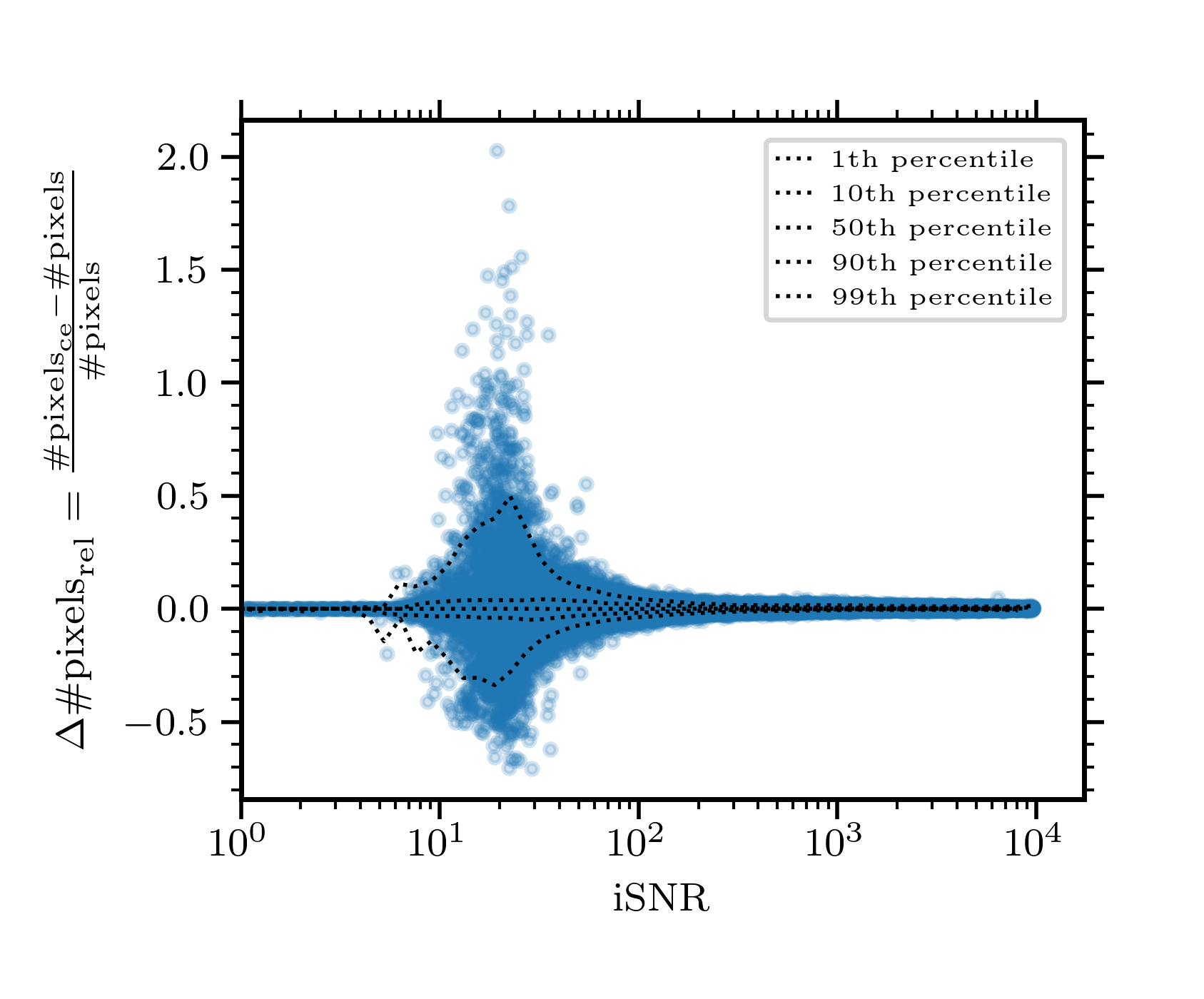}
		\caption{Relative change in number of pixels}
		\label{fig:pixel_rel_vs_snr_sub}
	\end{subfigure}
	\begin{subfigure}{0.5\textwidth}
		\centering
		\includegraphics[width=\textwidth]{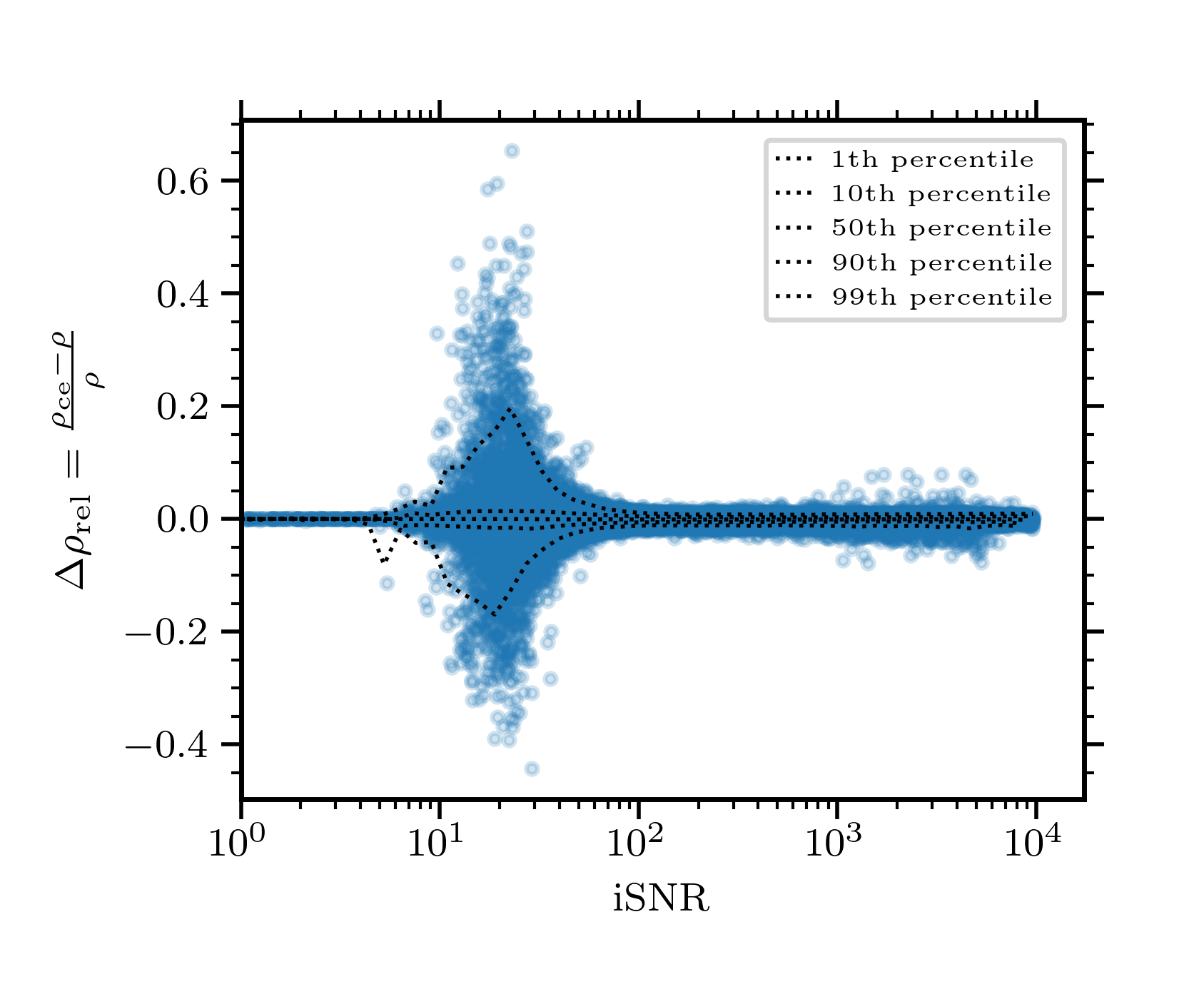}
		\caption{Relative change in the coherent network \gls{snr} $\rho$}
		\label{fig:rho_vs_snr_sub}
	\end{subfigure}%
	\begin{subfigure}{0.5\textwidth}
		\centering
		\includegraphics[width=\textwidth]{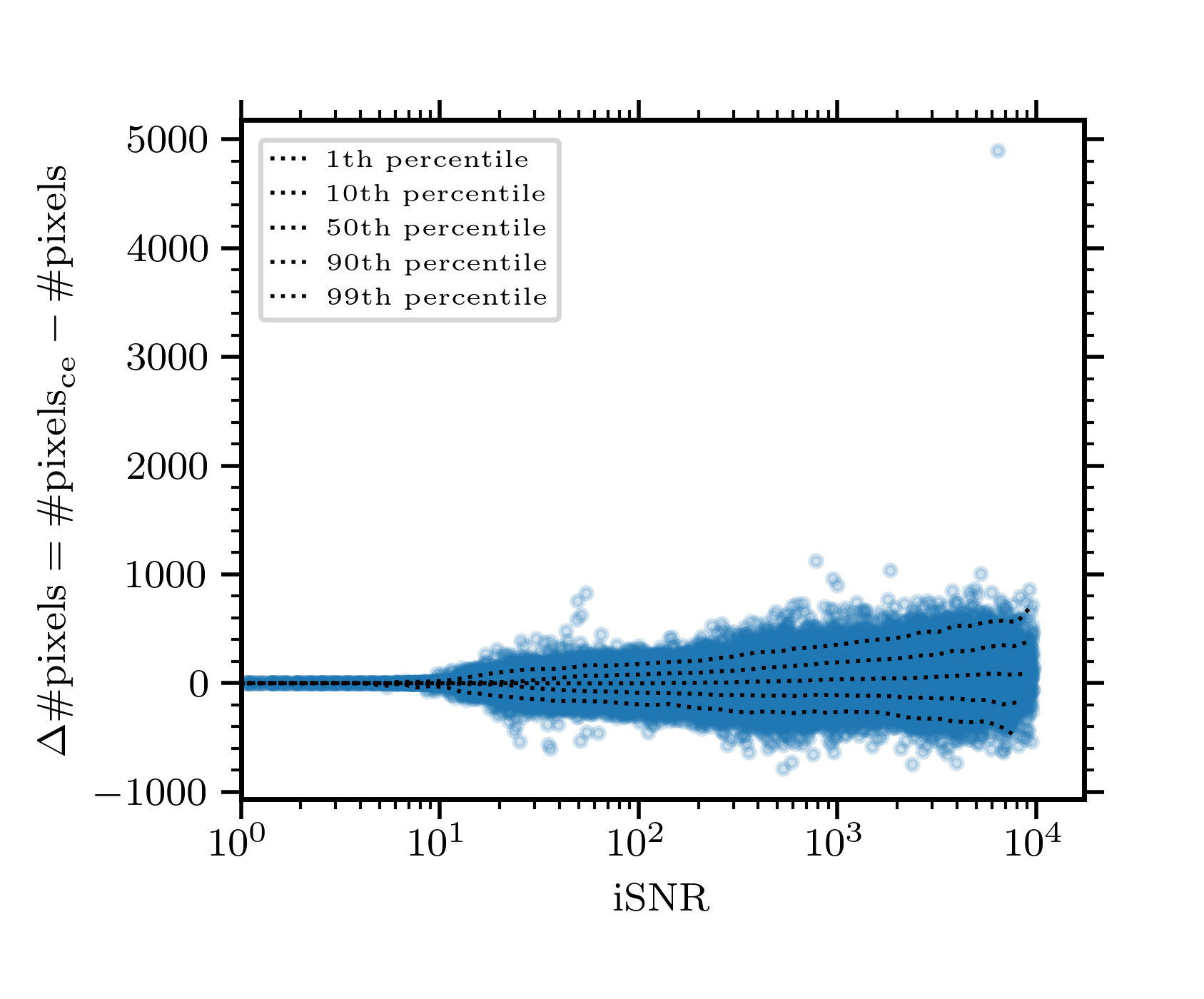}
		\caption{Absolute change in the number of pixels}
		\label{fig:pixel_abs_vs_snr_sub}
	\end{subfigure}
	\caption{Effect of calibration errors on the detection statistics of \gls{cwb} for the numerically simulated \gls{ccsn} waveforms listed in Table~\ref{tab:d90_comparison}. Each marker indicates an injection that was recovered by both the \gls{ce} and no-\gls{ce} analyses. The x-axis shows the injected network \gls{snr} without calibration errors.}
	\label{fig:combined_figures}
\end{figure}

The results of the run with numerical simulations of \gls{ccsn} waveforms are shown in figure~\ref{fig:combined_figures}.
Each marker indicates an injection that was recovered by both the \gls{ce} and no-\gls{ce} analyses.
The x-axis shows the injected network \gls{snr} without calibration errors.
For low \gls{snr} injections, the effect of calibration errors is negligible because all triggers are noise dominated and the noise in unaffected by calibration errors.
In fact, some of the triggers with the lowest \gls{snr} do not correspond to a reconstructed event but rather to a noise fluctuation that is close in time to the injection and is therefore matched to the injection.
This is possible because no post-production cuts are applied to the generated triggers.
This is necessary to get a complete picture of the effect of calibration errors on the detection statistics without biasing the results by removing triggers on the edge of the selection threshold.
As the \gls{snr} increases, the effect of calibration errors becomes more significant because the signal starts to dominate the detection statistics and the effect of calibration errors scales with the signal strength.

Figure~\ref{fig:cc_vs_snr_sub} shows the absolute change in correlation coefficient $cc$ as a function of the injected network \gls{snr}.
For medium \gls{snr} injections, the change in $cc$ can be significant, up to $0.2$ in absolute value.
For high \gls{snr} injections, the change in $cc$ is small, consistent with the scaling derived in equation~\ref{eq:network_correlation_coefficient_ce}.
The \gls{snr} range in which the effect of calibration errors is most significant is consistent with the detection threshold of \gls{cwb} for \gls{ccsn} waveforms, which is around $\mathrm{iSNR} \approx 10$.
This is expected as the signal level has to be comparable to the noise level for the effect of calibration errors to be significant, which is also the regime in which the detection threshold is located.
Note that the $10\%$ and $90\%$ percentiles of the absolute change in $cc$ remains below $0.1$ for all \gls{snr} values.
Additionally, the distribution of the change in $cc$ is approximately symmetric around zero for all \gls{snr} values.
Hence, the effect of calibration uncertainties on the correlation coefficient is an additional source of uncertainty but not a systematic bias.
This is a direct consequence of the first order effect of calibration errors on the detection statistics being linear in the calibration error and the calibration uncertainty being close to symmetric around zero.
If the calibration uncertainty were to be significantly skewed, then a systematic bias in the detection statistics could be observed because the expectation value over the calibration uncertainty in the equations in section~\ref{sec:first_order_impact} would no longer be zero.

The distribution of the relative change in the coherent network \gls{snr} $\rho$ in figure~\ref{fig:rho_vs_snr_sub} shows a similar trend as the correlation coefficient, with a significant effect for medium \gls{snr} injections and a suppressed effect for high \gls{snr} injections.
This is not consistent with the scaling derived in equations~\ref{eq:coherent_network_snr_ce_low_chi2} and \ref{eq:coherent_network_snr_ce_high_chi2} which predict an increase for low \gls{snr} and an asymptotic value for high \gls{snr}.

However, there is an another effect that is not captured by the first order calculations in section~\ref{sec:first_order_impact} which is the change in the number of pixels selected in the event.
Recall that the expressions in sections~\ref{sec:first_order_impact} are derived under the assumption of a fixed set of pixels, which is not the case in practice.
This appears to be caused by a large fluctuation in the number of detected pixels as shown in figure~\ref{fig:pixel_rel_vs_snr_sub}.
The distribution of the relative change in the number of detected pixels shown in figure~\ref{fig:pixel_rel_vs_snr_sub} is near-identical to the distribution of the relative change in the coherent network \gls{snr}.
This indicates that the change in the number of pixels is the main driver of the change in the coherent network \gls{snr}.

This variation in the number of pixels is caused by the selection threshold and clustering applied to the pixels in \texttt{cWB}.
For medium \gls{snr} injections, many pixels are close to the selection threshold.
Therefore a small change in the pixel \gls{snr} or coherence due to calibration errors can therefore lead to a large change in the number of selected pixels.
This secondary effect has been left out of most discussions about the impact of calibration errors on \gls{gw} detection but appears to be the dominant effect for unmodelled searches.
For high \gls{snr} injections, the absolute change in the number of pixels increases with \gls{snr} (figure~\ref{fig:pixel_abs_vs_snr_sub}) but the relative change decreases (figure~\ref{fig:pixel_rel_vs_snr_sub}).
Given that the affected pixels are mostly low-\gls{snr} pixels that are close to the selection threshold, the change in the number of pixels loses significance and the asymptotic behavior predicted by the first order calculations is recovered.

\begin{figure}
	\centering
	\begin{subfigure}{0.5\textwidth}
		\centering
		\includegraphics[width=\textwidth]{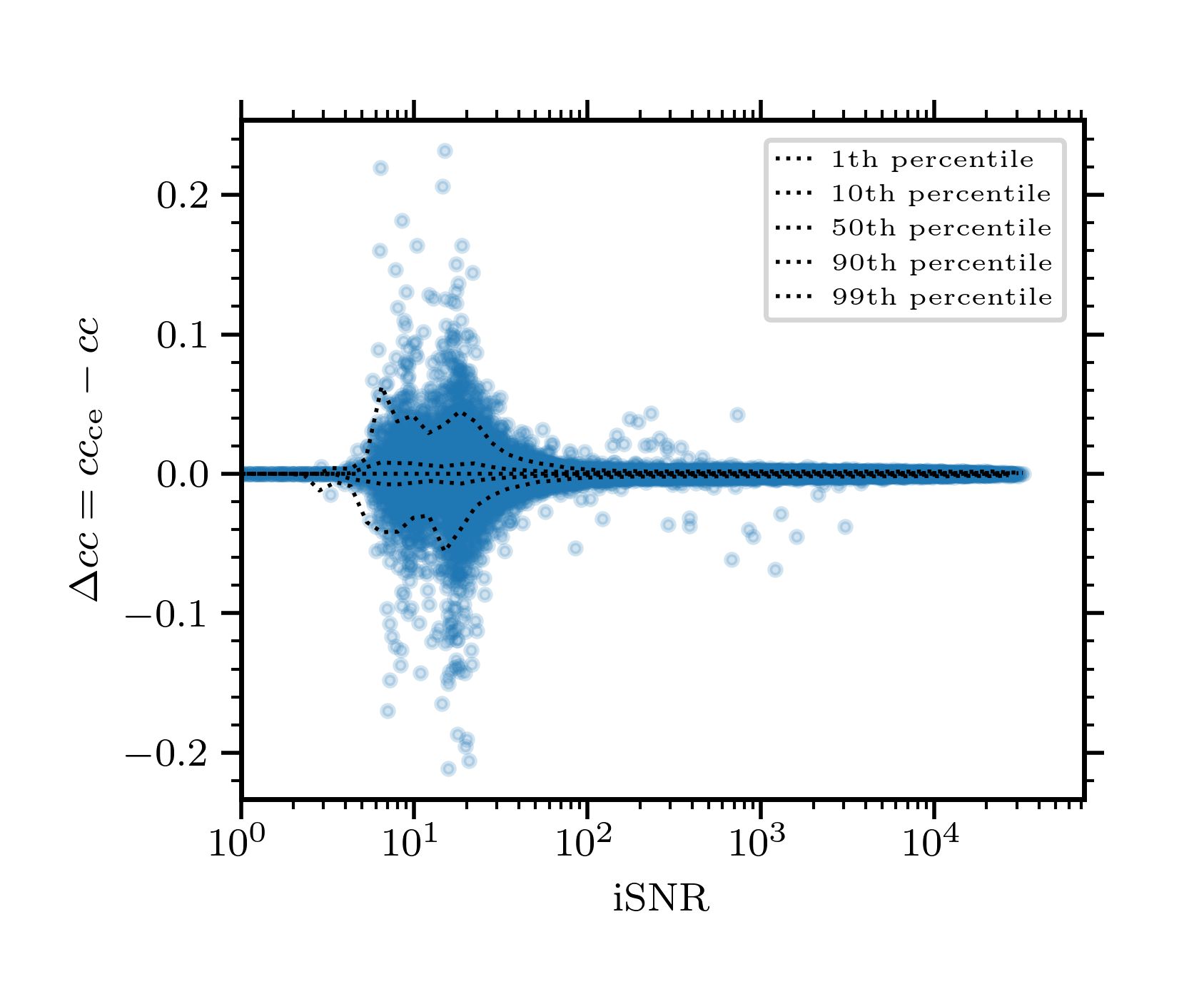}
		\caption{Absolute change in correlation coefficient $cc$}
	\end{subfigure}%
	\begin{subfigure}{0.5\textwidth}
		\centering
		\includegraphics[width=\textwidth]{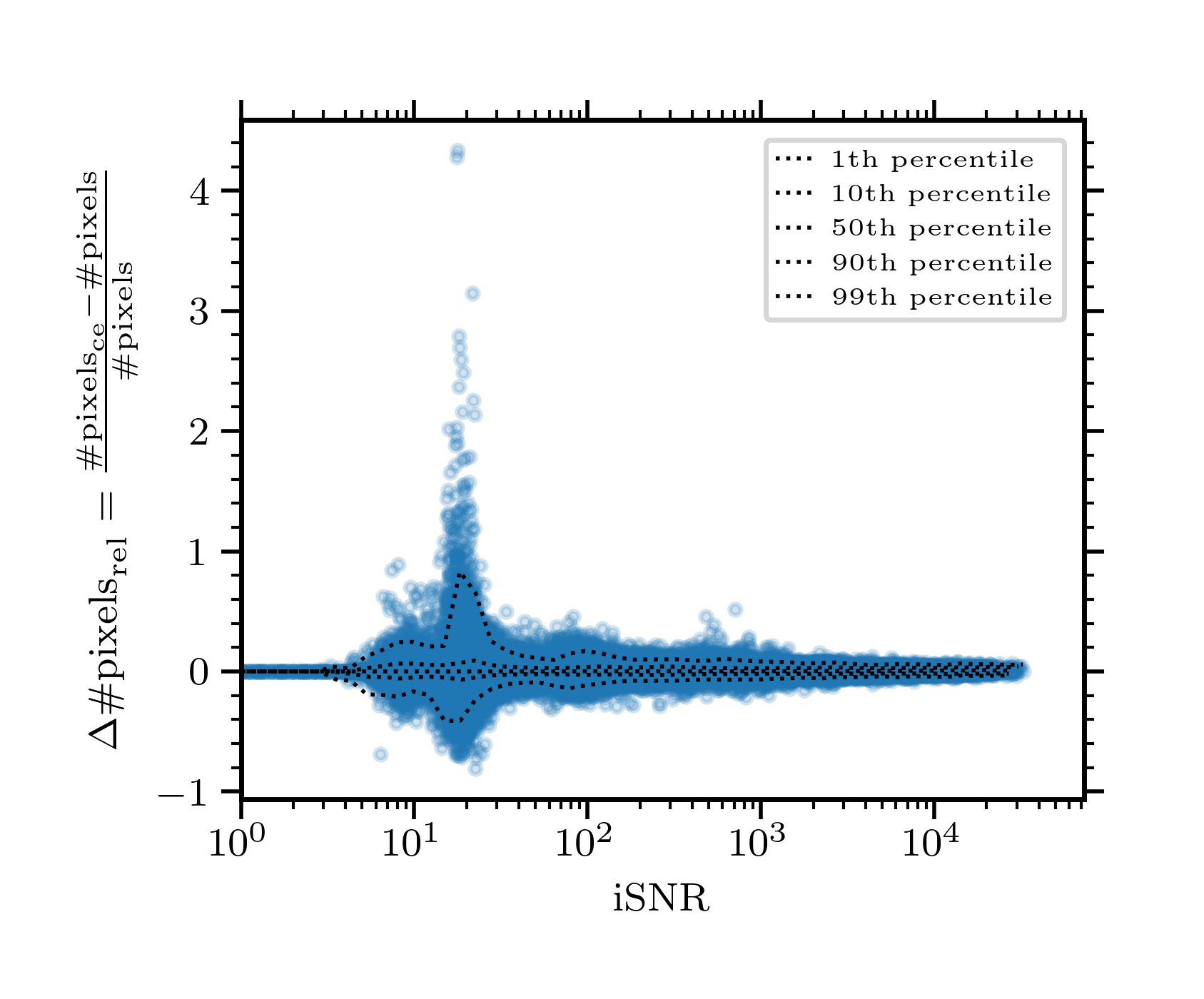}
		\caption{Relative change in number of pixels}
		\label{fig:pixel_rel_vs_snr_set2b}
	\end{subfigure}
	\begin{subfigure}{0.5\textwidth}
		\centering
		\includegraphics[width=\textwidth]{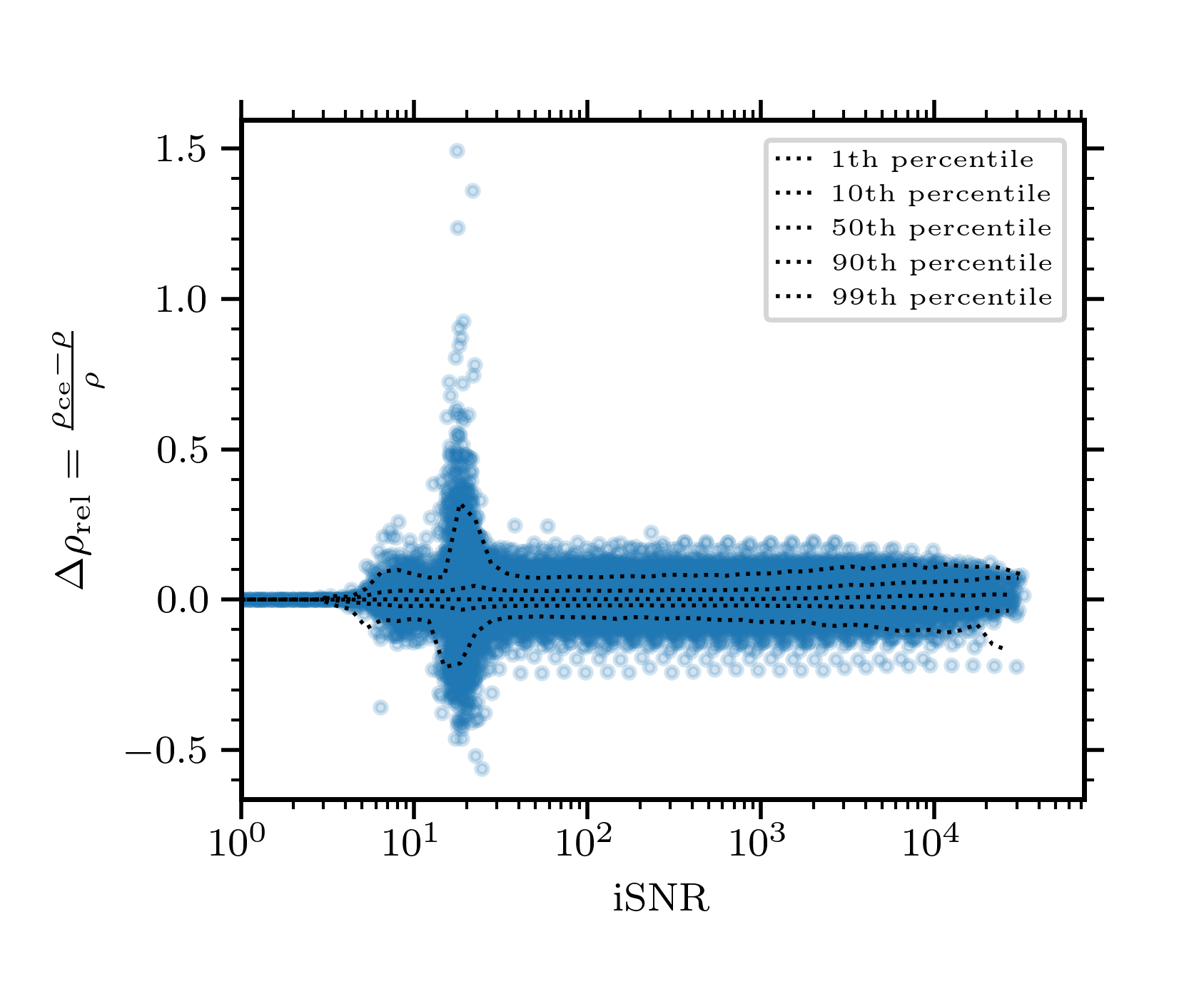}
		\caption{Relative change in the coherent network \gls{snr} $\rho$}
		\label{fig:rho_vs_snr_set2b}
	\end{subfigure}%
	\begin{subfigure}{0.5\textwidth}
		\centering
		\includegraphics[width=\textwidth]{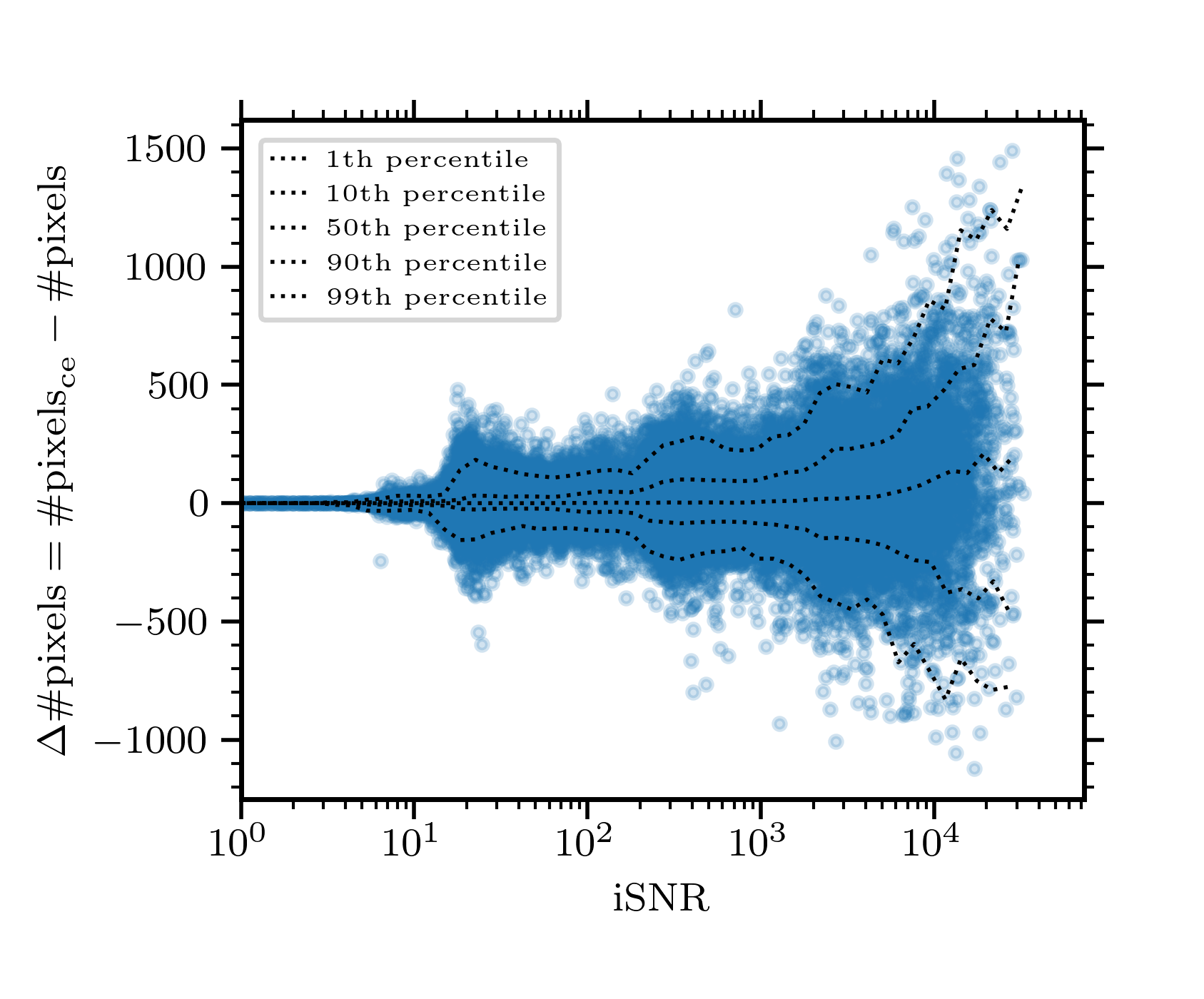}
		\caption{Absolute change in the number of pixels}
	\end{subfigure}
	\caption{Effect of calibration errors on the detection statistics of \gls{cwb} for a series of sine-Gaussian waveforms. Each marker indicates an injection that was recovered by both the \gls{ce} and no \gls{ce} analyses. The x-axis shows the injected network \gls{snr} without calibration errors.}
	\label{fig:combined_figures_set2b}
\end{figure}

The results of the simulations with sine-Gaussian waveforms are shown in figure~\ref{fig:combined_figures_set2b}.
The general trends are comparable to the simulations with \gls{ccsn} waveforms, with a significant effect of calibration errors for medium \gls{snr} injections and a suppressed effect for high \gls{snr} injections.
The variation in the number of pixels in figure~\ref{fig:pixel_rel_vs_snr_set2b} is more pronounced for sine-Gaussian waveforms, which could be caused by the more compact time-frequency representation of these waveforms.
This could also be the cause of the large asymptotic value of the relative change in the coherent network \gls{snr} in figure~\ref{fig:rho_vs_snr_set2b} for sine-Gaussian waveforms.
Note that some individual injections with a fixed calibration error for different scale factors are clearly visible and prove the expected asymptotic behavior for high \gls{snr} injections.
Additionally, the large variations in detection statistics occur over a narrower \gls{snr} range for sine-Gaussian waveforms.
This is likely caused by the morphology of the waveform comparable for all injections, while for numerically simulated \gls{ccsn} waveforms the morphology can vary significantly between injections.
Finally, the correlation coefficient has a few outliers at high \gls{snr} that do not follow the expected trend of suppressed calibration effects.
These outliers are characterized by a coherent \gls{snr} $\rho_{cwb} > \mathrm{iSNR}$ and an unusually low coherence $cc \approx 0.7-0.95$ for $\mathrm{iSNR} > 100$.
This indicates that these triggers might be affected by non-stationarities in the noise.

\section{Detection efficiency and upper limits}
\label{sec:impact_detection_efficiencies}

While the impact of calibration errors on the detection statistics gives great insight into the underlying mechanisms of how calibration errors affect the search, the main scientific concern is the impact on the number of detected events and the reconstructed waveform and parameters.
The performance of a search pipeline for a given signal class is typically quantified by the detection efficiency as a function of distance $P(\text{detection} | d)$, which is used to set upper limits on the rate of events.
In practice, this probability is estimated as the number of detected injections divided by the total number of injections at a given distance using the simulation procedure outlined in section~\ref{sec:simulation_setup}.
However, it is instructive to rewrite this probability in terms of the underlying distributions that influence the detection efficiency.

First, the detection efficiency can be written as a marginalisation over all possible detection statistics $\boldsymbol{x}$ that can be produced by a signal at a given distance $d$ as
\begin{align}
	P(\text{detection} | d) & = \int d\boldsymbol{x} \; p(\text{detection}, \boldsymbol{x} | d)                                            \\
	                        & = \int d\boldsymbol{x} \; P(\text{detection} | \boldsymbol{x}, d) p(\boldsymbol{x} | d)                      \\
	                        & = \int d\boldsymbol{x} \; \underbrace{\mathcal{I}(\boldsymbol{x})}_{\text{Classifier}} p(\boldsymbol{x} | d)
	\text{,}
\end{align}
where $P(\text{detection} | \boldsymbol{x}, d)$ is either zero or one and corresponds to a deterministic classifier (\gls{xgboost}).
Therefore, it replaced with an indicator function $\mathcal{I}(\boldsymbol{x})$ that is equal to one if a trigger with detection statistics $\boldsymbol{x}$ is classified as a signal and zero otherwise.
The distribution $p(\boldsymbol{x} | d)$ captures the variability in the detection statistics for signals at a given distance $d$ due to noise fluctuations, sky location, source orientation and calibration errors.
This can be made explicit by writing $p(\boldsymbol{x} | d)$ as a marginalisation over the noise $\boldsymbol{n}$, signal parameters $\boldsymbol{\theta}$ and calibration errors $\delta \tilde{\boldsymbol{C}}(f)$ as
\begin{equation}
	p(\boldsymbol{x} | d) = \int d\boldsymbol{n} \; d\boldsymbol{\theta} \; d\delta \tilde{\boldsymbol{C}}(f) \; p(\boldsymbol{x} | \boldsymbol{n}, \boldsymbol{\theta}, \delta \tilde{\boldsymbol{C}}(f), d) p(\boldsymbol{n}) p(\boldsymbol{\theta}) p(\delta \tilde{\boldsymbol{C}}(f))
	\text{.}
	\label{eq:detection_statistics_marginalisation}
\end{equation}

The calculations and results in section~\ref{sec:first_order_impact} show that the distribution of the detection statistics is merely broadened by the presence of calibration errors, as conjectured in the literature~\cite{2022PhRvD.105h2002E}, but it is not shifted if the distribution of calibration errors is symmetric.
Thus, if the distribution of detection statistics is located in region that is classified as signal, this broadening will lead to a decrease in the detection efficiency because some of the signals will now be classified as noise.
Conversely, if the distribution of detection statistics is located in region that is classified as noise, this broadening will lead to an increase in the detection efficiency because some of the signals will now be classified as signal.
Therefore, the distance at which the detection efficiency reaches 10\% ($d_{10\%}$) should increase, while the distance at which the detection efficiency reaches 90\% ($d_{90\%}$) should decrease, i.e. the detection efficiency curve is expected to be broadened by calibration errors.
Tables~\ref{tab:d90_comparison} and \ref{tab:d10_comparison} show the $d_{90\%}$ and $d_{10\%}$ values for a selection of \gls{ccsn} waveforms with and without calibration errors applied.
The results do not show the predicted broadening of the detection efficiency curve, with the differences in $d_{90\%}$ and $d_{10\%}$ being less than 1\% for all waveforms.

Most likely, the variation introduced by calibration errors is negligible compared to the variation introduced by the other factors in equation~\ref{eq:detection_statistics_marginalisation}.
Indeed, for a network of two detectors, the variation in \gls{snr} between different sky locations and source orientations can easily be a factor of two, which is much larger than the typical variations in the detection statistics due to calibration errors.
However, if the calibration errors were skewed, systematic bias in the detection statistics would occur and therefore a systematic increase or decrease in the detection efficiency.
Additionally, second order effects effects of calibration errors could potentially lead to a systematic bias in the detection statistics at the percent level if the calibration errors are $\sim 10\%$ but this is not observed in our results.
The remaining small differences in $d_{90\%}$ and $d_{10\%}$ are likely caused by statistical fluctuations due to the finite number of injections, which is supported by the fact that the differences are not consistent across waveforms and do not show a clear trend.

\begin{table}
	\centering
	\begin{tabular}{lccc}
		\hline
		Waveform     & d90 (CE) & d90 (No CE) & Diff
		\\
		\hline
		s11          & 0.945    & 0.947       & -0.21\% \\
		SFHx         & 7.57     & 7.61        & -0.53\% \\
		D15-3D       & 3.13     & 3.13        & 0.00\%  \\
		mesa20       & 0.892    & 0.889       & 0.34\%  \\
		mesa20\_pert & 1.09     & 1.08        & 0.93\%  \\
		s18\_3d      & 5.79     & 5.80        & -0.17\% \\
		s3.5\_pns    & 2.99     & 3.00        & -0.33\% \\
		s13          & 0.615    & 0.618       & -0.49\% \\
		s25          & 6.17     & 6.18        & -0.16\% \\
		NR           & 6.96     & 6.95        & 0.14\%  \\
		\hline
	\end{tabular}
	\caption{90\% detection efficiencies in kpc for a selection of \gls{ccsn} waveforms with and without calibration errors (CE) applied.}
	\label{tab:d90_comparison}
\end{table}

\begin{table}
	\centering
	\begin{tabular}{lccc}
		\hline
		Waveform     & d10 (CE) & d10 (No CE) & Diff
		\\
		\hline
		s11          & 2.07     & 2.07        & 0.00\%  \\
		SFHx         & 22.8     & 22.9        & -0.44\% \\
		D15-3D       & 6.36     & 6.37        & -0.16\% \\
		mesa20       & 1.97     & 1.97        & 0.00\%  \\
		mesa20\_pert & 2.59     & 2.64        & -1.89\% \\
		s18\_3d      & 11.7     & 11.7        & 0.00\%  \\
		s3.5\_pns    & 5.86     & 5.87        & -0.17\% \\
		s13          & 1.54     & 1.54        & 0.00\%  \\
		s25          & 13.3     & 13.3        & 0.00\%  \\
		NR           & 15.8     & 15.8        & 0.00\%  \\
		\hline
	\end{tabular}
	\caption{10\% detection efficiencies in kpc for a selection of \gls{ccsn} waveforms with and without calibration errors (CE) applied.}
	\label{tab:d10_comparison}
\end{table}

Similar arguments can be made for the detection efficiency as a function of the root-sum-square strain amplitude
\begin{equation}
	h_{rss} = \sqrt{\int_{-\infty}^{\infty} dt \; (h_{+}(t)^2 + h_{\times}(t)^2)}
	\text{.}
\end{equation}
The value of the injected $h_{rss}$ at which the detection efficiency reaches 90\% ($h_{rss, 90\%}$) is used to set upper limits on the explosion energy in \gls{gw} for \gls{ccsn} searches~\cite{2025ApJ...985..183A}
\begin{equation}
	E_{\mathrm{GW}} = \frac{2}{5}\frac{{\pi }^{2}c^{3}}{G} \sqrt{\frac{\pi}{2}} \tau D^2 f_{0}^{2} h_{\mathrm{rss}}^{2}
	\label{eq:gw_energy}
\end{equation}
Where $\tau$ is the duration of the signal, $D$ is the distance to the source and $f_{0}$ is the central frequency of the signal.
The uncertainty on the estimated detectable explosion energy is therefore
\begin{equation}
	\frac{\delta E_{\mathrm{GW}}}{E_{\mathrm{GW}}} = 2 \sqrt{ \left( \frac{\delta D}{D} \right)^{2} + \left( \frac{\delta h_{\mathrm{rss}}}{h_{\mathrm{rss}}} \right)^{2} }
	\text{.}
	\label{eq:gw_energy_uncertainty}
\end{equation}
Given that the uncertainty on $h_{rss}$ is less than 1\%, the uncertainty on the estimated explosion energy is dominated by the uncertainty on the distance to the host galaxy, which is of the order of 10\% for SN 2023ixf~\cite{2025ApJ...985..183A}.

\section{Conclusion}

In this work, we investigated the impact of realistic, frequency‑dependent calibration errors on the performance of unmodelled gravitational-wave burst searches, with a particular focus on broadband signals from core-collapse supernovae.
Unlike previous studies, which relied on simplified calibration modes such as global amplitude rescaling or time jitter, we implemented a new calibration error plugin for \texttt{cWB} that allows physically motivated and frequency‑dependent distortions to be applied directly to injected waveforms.
This capability enables, for the first time, a systematic study of how the full calibration uncertainty budget affects the detection statistics used in current LVK burst analyses.

First‑order analytical estimates were developed to clarify how calibration errors influence different \gls{cwb} detection statistics as a function of signal strength.
These calculations predict that the relative impact on the coherent network \gls{snr} should increase with signal strength for low-\gls{snr} events and approach an asymptotic level for high‑\gls{snr} signals.
Conversely, the effect on the correlation coefficient is expected to peak for medium-\gls{snr} events and diminish for high-\gls{snr} signals.
Large-scale injection campaigns confirmed that calibration errors do modify \texttt{cWB}'s detection statistics, but they also revealed that the dominant contribution arises from changes in the number of pixels selected in an event, a secondary effect not captured by the first‑order model.

Despite these variations in individual detection statistics, the impact on astrophysical conclusions is negligible for core-collapse supernova searches.
Detection efficiencies as a function of distance and root‑sum‑square strain differ by less than one percent across all tested waveforms.
This behaviour can be understood from the fact that statistical fluctuations due to sky position, source orientation, and noise realisation are substantially larger than any variation introduced by calibration errors.
This is fully consistent with the first‑order predictions, which indicate a symmetric broadening of detection‑statistic distributions in the absence of systematic calibration biases.
As a result, explosion‑energy upper limits remain dominated by astrophysical uncertainties, most notably the distance to the host galaxy, rather than by calibration uncertainty.

Overall, our results demonstrate that although calibration errors can alter the detailed morphology of broadband burst signals, their effect on the detectability of core-collapse supernovae with \texttt{cWB} is minimal at current detector sensitivity.
For future detectors such as the Einstein Telescope and Cosmic Explorer, which will have significantly improved sensitivity, the impact of calibration errors will need to be reassessed.
We expect that the main effect on the detection efficiency will still be the variation in pixel selection around the detection threshold rather than a systematic bias in the detection statistics.
However, for \gls{ccsn} signals the improved sensitivity at low frequencies might turn the \gls{gw} memory effect into the dominant contribution to the detection statistics.
As a result, the calibration errors in the narrow \SIrange{2}{10}{\hertz} band around the memory effect will likely have a more significant impact on the detection efficiency than the \SIrange{20}{2000}{\hertz} band that is currently the most relevant for \gls{ccsn} searches.

Finally, the newly developed calibration error plugin provides a robust and flexible framework for incorporating calibration uncertainties into future burst analyses and will continue to be valuable as detector calibration and search methodology advance.

%
%

\ack{
	This material is based upon work supported by NSF's LIGO Laboratory which is a major facility fully funded by the National Science Foundation.
	The authors are grateful for computational resources provided by the LIGO Laboratory and supported by National Science Foundation Grants PHY-0757058 and PHY-0823459.
}

\funding{
	M.W. is supported by the \gls{fwo} through Grant No. 11POK24N.
	M.S. acknowledges Polish National Science Centre Grants No. UMO-2023/49/B/ST9/02777 and No. UMO-2024/03/1/ST9/00005, and the Polish National Agency for Academic Exchange within Polish Returns Programme Grant No. BPN/PPO/2023/1/00019
	M.Z. is supported by the National Science Foundation Gravitational Physics Experimental and Data Analysis Program through awards PHY-2110555 and PHY-2405227.
	T.G.F. Li is partially supported by the \gls{fwo} through Grant No. I002123N.
}

\data{The strain data used in this paper is publicly available through the Gravitational Wave Open Science Center:~\url{https://gwosc.org/data/}.}

\appendix

\section{Incoherent energy for two detector paradox}
\label{sec:incoherent_energy_derivation}

In the case of two non co-aligned detectors, the projection matrix is defined as
\begin{equation}
	\mathbf{P}_{sig}[i] = \frac{\boldsymbol{w}[i] \boldsymbol{w}[i]^H}{\boldsymbol{w}[i]^H \boldsymbol{w}[i]} \text{,}
\end{equation}
i.e. the projection onto the one-dimensional subspace spanned by $\boldsymbol{w}[i]$.
As a result, the incoherent energy is given by
\begin{align}
	E_{i} & = \sum_{i \in C } \sum_{m} (w_m[i])^H (P_{sig}[i])_{mm} w_m[i] \nonumber                                               \\
	      & = \sum_{i \in C } \sum_{m} (w_m[i])^H \frac{(w_m[i])^H w_m[i]}{\boldsymbol{w}[i]^H \boldsymbol{w}[i]} w_m[i] \nonumber \\
	      & = \sum_{i \in C } \frac{\sum_{m} \left( (w_m[i])^H w_m[i] \right)^2}{\boldsymbol{w}[i]^H \boldsymbol{w}[i]}
\end{align}
This can be expanded to first order in a variation of the elements of the data vector $\mathbf{w}[i]$.
Because it is a real function expressed in terms of the complex variables $w_m[i]$ and $(w_m[i])^H$, the variation can be expressed as
\begin{equation}
	\delta E_{i} = \sum_{i \in C } \sum_{m} 2 \Re \left[ \frac{\partial E_{i}}{\partial w_m[i]} \delta w_m[i] \right] + \mathcal{O}(\delta \boldsymbol{w}^2)
	\text{,}
\end{equation}
where $\frac{\partial}{\partial w_m[i]}$ denotes the Wirtinger derivative with respect to $w_m[i]$.
Calculating the Wirtinger derivative gives
\begin{align}
	\frac{\partial E_{i}}{\partial w_m[i]} & = \frac{2 (w_m[i])^H w_m[i] (w_m[i])^H}{\boldsymbol{w}[i]^H \boldsymbol{w}[i]} - \frac{\sum_{m} \left( (w_m[i])^H w_m[i] \right)^2 (w_m[i])^H}{(\boldsymbol{w}[i]^H \boldsymbol{w}[i])^2} \nonumber \\
	                                       & = \frac{(w_m[i])^H}{\boldsymbol{w}[i]^H \boldsymbol{w}[i]} \left( 2 (w_m[i])^H w_m[i] - E_i \right)
	\text{.}
\end{align}
Therefore, the first order variation of the incoherent energy is given by
\begin{equation}
	\delta E_{i} = \sum_{i \in C } 2 \Re \left[ (\boldsymbol{w}[i])^H \mathbf{D}[i] \delta \boldsymbol{\xi}[i] \right] + \mathcal{O}(\delta \boldsymbol{\xi}^2)
	\text{,}
\end{equation}
where $\mathbf{D}[i]$ is a diagonal matrix with elements
\begin{equation}
	(\mathbf{D}[i])_{mm} = \frac{2 (w_m[i])^{H} w_m[i] - E_i}{(\boldsymbol{w}[i])^{H}\boldsymbol{w}[i]}
	\label{eqn:D_matrix_paradox}
\end{equation}
and the first order variation $\delta \boldsymbol{w}[i]$ is replaced by $\delta \boldsymbol{\xi}[i]$ because the variation of the data vector is caused by the calibration errors.

\section{Artefacts from Calibration Errors}

Applying spectral modifications to discrete non-periodic signals is a common problem in digital signal processing.
\Gls{ola}, overlap-save and \gls{wola} are standard algorithms to tackle this problem.
The general procedure consists of a multiplication of the time domain signal $x[k]$ with a window function $w[k]$ that has a finite length $M$.
The signal is then extended with zero padding to size $N$ before performing a \gls{dft}.
Next, the frequency modification is performed by multiplying with a frequency response $H[n]$.
Finally, the signal is transformed back to the time domain with an \gls{idft}.

The window is then shifted by $D$ samples and the procedure is repeated.
The difference between different algorithms is how the signal is reconstructed from the outputs of separate windows.
As the name states, \gls{ola} adds the overlapping outputs together.
Any window that satisfies the \gls{cola} constraint
\begin{equation}
	\sum_{l=-\infty}^{+\infty} w[k - l D] = 1 \quad \forall \; k \in \mathds{Z}
\end{equation}
can be used with this method but a rectangular window is most common because of its simplicity.
The multiplication in the frequency domain corresponds to a circular convolution in the time domain.
This causes artefacts, unless the \gls{dft} size $N$ is greater than, or equal to the length of the linear convolution of the windowed signal with the impulse response $h[k]$ corresponding to $H[n]$
\begin{equation}
	N \geq M + L - 1 \text{,}
\end{equation}
where $L$ is the length of $h[k]$.
Overlap-save uses a rectangular window and discards the first $L-1$ and last $N-M$ samples because they contain edge artefacts.
The shift parameter $D$ is chosen to be $D = M - L + 1$ to seamlessly stitch the output together.
Finally, \gls{wola} applies an additional window after the \gls{idft} which leads to a modified \gls{cola} condition
\begin{equation}
	\sum_{l=-\infty}^{+\infty} w^2[k - l D] = 1 \quad \forall \; k \in \mathds{Z} \text{.}
\end{equation}
Note that \gls{wola} is only approximately equal to a time-domain convolution but the modified \gls{cola} condition ensures that in the absence of any spectral modification the signal is perfectly reconstructed.
Additionally, the zero padding is usually removed because \gls{wola} is mostly used when the spectral modification does not correspond to a \gls{fir} filter.

Calibration errors are characterised in the frequency domain with a transfer function
\begin{equation}
	\delta \Tilde{R} = \frac{\Tilde{R}^{(meas)}}{\Tilde{R}^{(model)}} \text{,}
\end{equation}
where $\Tilde{R}^{(meas)}$ is the measured response function.
Hence, there is no guarantee that this frequency response corresponds to an \gls{fir} filter, and \gls{wola} could be a suitable method to perform the spectral modification.
However, the \gls{idft} matrix the hermitian conjugate of the \gls{dft} matrix $\mathcal{F}^{-1} = \mathcal{F}^{\mathbf{H}}$, and provides a one-to-one mapping from the discretised calibration errors $\delta \mathbf{R}$ to its time domain representation $\delta \mathbf{r}$
\begin{equation}
	\begin{aligned}
		\delta \mathbf{r} & = \mathcal{F}^{\mathbf{H}} \; \delta \mathbf{R} \\
		                  & = \mathcal{F}^{\text{T}} \; \delta \mathbf{R}^* \\
		                  & = \mathcal{F} \; \delta \mathbf{R}^*
	\end{aligned}
	\text{,}
	\label{eqn:deltar}
\end{equation}
where $*$ denotes the complex conjugate.
From equation~\ref{eqn:deltar}, it follows that $\delta r[k] = 0$ around $k = N / 2$ when the $\delta \Tilde{R}$ varies slowly with the frequency.
Because of periodicity in $r[k]$, the non-zero part below and above $N / 2$ can be interpreted as the causal and anti-causal part of an \gls{fir} filter, respectively.

The anti-causal tail causes wrap-around artefacts in the \gls{ola} algorithm.
This can be avoided by adding zero-padding before the windowed signal.
Let $h[k]$ be a finite, non-causal impulse response
\begin{equation}
	h[k] = 0 \quad \forall \; k \geq L \; \text{and} \; k \leq L'
	\text{,}
\end{equation}
and $x[k]$ be a zero-padded windowed signal
\begin{equation}
	x[k] = 0 \quad \forall \; 0 \leq \; k < M' \; \text{and} \; M \leq\; k < N
	\text{.}
\end{equation}
Then the circular convolution will not introduce any artefacts as long as the anti-causal part of $h[k]$ is shorter than the zero padding at the start of $x[k]$, and the causal part is shorter than the zero padding at the end of the signal
\begin{gather}
	L'\geq 1 - M'  \\
	L \leq N - M + 1
	\text{.}
\end{gather}
Therefore, if $\delta \Tilde{R}$ is sufficiently smooth, it is always possible to add sufficient zero-padding such that the output is artefact-free.
For overlap-save it suffices to discard samples $M + L' + 1, \cdots , M-1$ and adjust the shift parameter accordingly.

\printbibliography

\end{document}

%% file: acronyms.tex
\newacronym{adc}{ADC}{analog-to-digital converter}
\newacronym{ce}{CE}{calibration error}
\newacronym{cola}{COLA}{constant overlap-add}
\newacronym{cbc}{CBC}{compact binary coalescence}
\newacronym[shortplural=CCSNe,longplural=core-collapse supernovae]{ccsn}{CCSN}{core-collapse supernova}
\newacronym{cwb}{\texttt{cWB}}{coherent WaveBurst}
\newacronym{dft}{DFT}{discrete Fourier transform}
\newacronym{darm}{DARM}{differential arm}
\newacronym{dac}{DAC}{digital-to-analog converter}
\newacronym{fir}{FIR}{finite impulse response}
\newacronym{fwo}{FWO}{Research Foundation - Flanders}
\newacronym{gpr}{GPR}{Gaussian process regression}
\newacronym{gw}{GW}{gravitational wave}
\newacronym{idft}{IDFT}{inverse discrete Fourier transform}
\newacronym{ligo}{LIGO}{Laser Interferometer Gravitational-Wave Observatory}
\newacronym{lvk}{LVK}{LIGO-Virgo-KAGRA}
\newacronym{mcmc}{MCMC}{Markov chain Monte Carlo}
\newacronym{ola}{OLA}{overlap-add}
\newacronym{pns}{PNS}{proto-neutron star}
\newacronym{psd}{PSD}{power spectral density}
\newacronym{sasi}{SASI}{standing accretion shock instability}
\newacronym{snr}{SNR}{signal-to-noise ratio}
\newacronym{wdm}{WDM}{Wilson-Daubechies-Meyer}
\newacronym{wola}{WOLA}{weighted overlap-add}
\newacronym{xgboost}{XGBoost}{extreme gradient boosting}
\newacronym{xp}{XP}{cross-power}